\documentstyle[11pt,aaspp4]{article}
\begin{document}
\newbox\grsign \setbox\grsign=\hbox{$>$} \newdimen\grdimen \grdimen=\ht\grsign
\newbox\simlessbox \newbox\simgreatbox
\setbox\simgreatbox=\hbox{\raise.5ex\hbox{$>$}\llap
     {\lower.5ex\hbox{$\sim$}}}\ht1=\grdimen\dp1=0pt
\setbox\simlessbox=\hbox{\raise.5ex\hbox{$<$}\llap
     {\lower.5ex\hbox{$\sim$}}}\ht2=\grdimen\dp2=0pt
\def\gtorder{\mathrel{\copy\simgreatbox}}
\def\ltorder{\mathrel{\copy\simlessbox}}
\def\simgreat{\mathrel{\copy\simgreatbox}}
\def\simless{\mathrel{\copy\simlessbox}}

\def\chaphead{}

\def\hut{Hubble type\ }
\def\vc{V$_{\rm C}$\ }
\def\mb{M$_{\rm B}$\ }
\def\av{A$_{\rm V}$\ }
\def\lamlam{$\lambda\lambda$}

\def\deg{$^\circ$}
\def\degrees{$^\circ$}
\def\Vlasov{collisionless Boltzmann\ }
\def\lsls{\ll}
\def\grgr{\gg}
\def\erf{\mathop{\rm erf}\nolimits} 
\def\eqv{\equiv}
\def\real{\Re e}
\def\imag{\Im m}
\def\ctrline#1{\centerline{#1}}
\def\spose#1{\hbox to 0pt{#1\hss}}
     
\def\={\overline}
\def\sections{\S}
\newcount\notenumber
\notenumber=1
\newcount\eqnumber
\eqnumber=1
\newcount\fignumber
\fignumber=1
\newbox\abstr
\newbox\figca     
\def\yyskip{\penalty-100\vskip6pt plus6pt minus4pt}
     
\def\numberpara{\yyskip\noindent}
     
\def\km{{\rm\,km}}
\def\kms{{\rm\ km\ s$^{-1}$}}
\def\kpc{{\rm\,kpc}}
\def\mpc{{\rm\,Mpc}}
\def\etal{{\it et al. }}
\def\eg{{\it e.g. }}
\def\ie{{\it i.e. }}
\def\cf{{\it cf. }}
\def\msun{{\rm\,M_\odot}}
\def\lsun{{\rm\,L_\odot}}
\def\rsun{{\rm\,R_\odot}}
\def\pc{{\rm\,pc}}
\def\cm{{\rm\,cm}}
\def\yr{{\rm\,yr}}
\def\au{{\rm\,AU}}
\def\AU{{\rm\,AU}}
\def\gm{{\rm\,g}}
\def\s{{\rmss}}
\def\dyne{{\rm\,dyne}}
     
\def\note#1{\footnote{$^{\the\notenumber}$}{#1}\global\advance\notenumber by 1}
\def\foot#1{\raise3pt\hbox{\eightrm \the\notenumber}
     \hfil\par\vskip3pt\hrule\vskip6pt
     \noindent\raise3pt\hbox{\eightrm \the\notenumber}
     #1\par\vskip6pt\hrule\vskip3pt\noindent\global\advance\notenumber by 1}
\def\propo{\propto}
\def\larrow{\leftarrow}
\def\rarrow{\rightarrow}
\def\sectionhead#1{\penalty-200\vskip24pt plus12pt minus6pt
        \centerline{\bbrm#1}\vskip6pt}
     
\def\Dt{\spose{\raise 1.5ex\hbox{\hskip3pt$\mathchar"201$}}}    
\def\dt{\spose{\raise 1.0ex\hbox{\hskip2pt$\mathchar"201$}}}    
\def\llangle{\langle\langle}
\def\rrangle{\rangle\rangle}
\def\ldotss{\ldots}
\def\del{\b\nabla}
     
\def\new{{\rm\chaphead\the\eqnumber}\global\advance\eqnumber by 1}
\def\ref#1{\advance\eqnumber by -#1 \chaphead\the\eqnumber
     \advance\eqnumber by #1 }
\def\last{\advance\eqnumber by -1 {\rm\chaphead\the\eqnumber}\advance
     \eqnumber by 1}
\def\eqnam#1{\xdef#1{\chaphead\the\eqnumber}}
     
\def\nfig{\chaphead\the\fignumber\global\advance\fignumber by 1}
\def\nfiga#1{\chaphead\the\fignumber{#1}\global\advance\fignumber by 1}
\def\rfig#1{\advance\fignumber by -#1 \chaphead\the\fignumber
     \advance\fignumber by #1}
\def\refindent{\par\noindent\parskip=3pt\hangindent=3pc\hangafter=1 }

\def\apj#1#2#3{\refindent#1,  {ApJ,\ }{\bf#2}, #3}
\def\apjsup#1#2#3{\refindent#1,  {ApJS,\ }{\bf#2}, #3}
\def\aasup#1#2#3{\refindent#1,  { A \& AS\ }{\bf#2}, #3}
\def\aas#1#2#3{\refindent#1,  { Bull. Am. Astr. Soc.,\ }{\bf#2}, #3}
\def\apjlett#1#2#3{\refindent#1,  { ApJL,\  }{\bf#2}, #3}
\def\mn#1#2#3{\refindent#1,  { MNRAS,\ }{\bf#2}, #3}
\def\mnras#1#2#3{\refindent#1,  { M.N.R.A.S., }{\bf#2}, #3}
\def\annrev#1#2#3{\refindent#1, { ARA \& A,\ }
{\bf2}, #3}
\def\aj#1#2#3{\refindent#1,  { AJ,\  }{\bf#2}, #3}
\def\phrev#1#2#3{\refindent#1, { Phys. Rev.,}{\bf#2}, #3}
\def\aa#1#2#3{\refindent#1,  { A \& A,\ }{\bf#2}, #3}
\def\nature#1#2#3{\refindent#1,  { Nature,\ }{\bf#2}, #3}
\def\icarus#1#2#3{\refindent#1,  { Icarus, }{\bf#2}, #3}
\def\pasp#1#2#3{\refindent#1,  { PASP,\ }{\bf#2}, #3}
\def\appopt#1#2#3{\refindent#1,  { App. Optics,\  }{\bf#2}, #3}
\def\spie#1#2#3{\refindent#1,  { Proc. of SPIE,\  }{\bf#2}, #3}
\def\opteng#1#2#3{\refindent#1,  { Opt. Eng.,\  }{\bf#2}, #3}
\def\refpaper#1#2#3#4{\refindent#1,  { #2 }{\bf#3}, #4}
\def\refbook#1{\refindent#1}
\def\science#1#2#3{\refindent#1, { Science, }{\bf#2}, #3}
     
\def\chapbegin#1#2{\eject\vskip36pt\par\noindent{\chapheadfont#1\hskip30pt
     #2}\vskip36pt}
\def\sectionbegin#1{\vskip30pt\par\noindent{\bf#1}\par\vskip15pt}
\def\subsectionbegin#1{\vskip20pt\par\noindent{\bf#1}\par\vskip12pt}
\def\topic#1{\vskip5pt\par\noindent{\topicfont#1}\ \ \ \ \ }
     
\def\ltsim{\mathrel{\spose{\lower 3pt\hbox{$\mathchar"218$}}
     \raise 2.0pt\hbox{$\mathchar"13C$}}}
\def\gtsim{\mathrel{\spose{\lower 3pt\hbox{$\mathchar"218$}}
     \raise 2.0pt\hbox{$\mathchar"13E$}}}
     
\def\sec{\hbox{$^s$\hskip-3pt .}}
\def\gg{\hbox{$>$\hskip-4pt $>$}}
\parskip=3pt
\def\gapprox{$_ >\atop{^\sim}$}     
\def\lapprox{$_ <\atop{^\sim}$}     
\def\apequal{\mathrel{\spose{\lower 1pt\hbox{$\mathchar"218$}}
     \raise 2.0pt\hbox{$\mathchar"218$}}}

 \def\oforder{$\sim$} \def\inv{$^{-1}$}
\def\>={$\geq$} \def\<={$\leq$} \def\ks{km s\inv} \def\kms{km s\inv}
\def\lith{$h$} \def\sig{$\sigma$} \def\sigp{$\sigma^{\prime}_r$}
\def\meanz{$\overline \upsilon$} \def\nc{$N_c$} \def\rc{$r_c$}
\def\twidle{$\sim$} \def\sigmar{$\sigma_r$}
\def\Mstar{{M}^*_R}
\def\Mdot{M_{\odot}}

\title{THE PROPERTIES OF POOR GROUPS OF GALAXIES:  III. THE GALAXY LUMINOSITY FUNCTION} 
\author{Ann I. Zabludoff\altaffilmark{1}}
\affil{UCO/Lick Observatory and Board of Astronomy and Astrophysics,} 
\centerline{University of California at Santa Cruz, Santa Cruz, CA 95064}
\centerline{aiz@ucolick.org}

\centerline{and}

\author{John S. Mulchaey}
\affil{The Observatories of the Carnegie Institution of Washington,}
\centerline {813 Santa Barbara St., Pasadena, CA 91101}
\centerline{mulchaey@pegasus.ociw.edu}
\bigskip
\centerline {Accepted for publication in {\it The Astrophysical Journal}}

\altaffiltext{1}{New Address:  University of Arizona, Steward Observatory,
Tucson, AZ 85721, E-mail:  azabludoff@as.arizona.edu}

\singlespace
\abstract{
The form of the galaxy luminosity function (GLF) in poor groups 
--- regions of intermediate galaxy density that are common 
environments for galaxies --- is not well understood.
Multi-object spectroscopy and
wide-field CCD imaging now allow us to measure the GLF of bound
group members directly ({\it i.e.}, 
without statistical background subtraction) and to compare
the group GLF with the GLF's of the field and of rich clusters.
We use R-band images in $1.5 \times 1.5$
degree$^2$ mosaics to obtain photometry for galaxies in the fields of
six nearby ($2800 < cz < 7700$ \ks) poor groups 
for which we have extensive spectroscopic data
(Zabludoff \& Mulchaey 1998), including 328 new galaxy velocities (this paper).
For the five groups with luminous X-ray halos,
the composite group GLF for group members with $-23 + 5$log $h < M_R <
-16 + 5$log $h$ and within projected radii of $\simless 0.4-0.6$\lith\inv\ Mpc
from the group center is fit adequately by a Schechter function with
$\Mstar = -21.6 \pm 0.4 + 5$log $h$ and $\alpha = -1.3 \pm 0.1$.

We also find that (1) the ratio of dwarfs 
($-17 + 5$log $h \geq M_R > -19 + 5$log $h$) to giants
($M_R \leq -19 + 5$log $h$) is significantly larger for the five
groups with luminous X-ray halos than for
the one marginally X-ray detected group,
(2) the composite GLF for the luminous X-ray groups is
consistent in shape with two measures
of the composite R-band GLF for rich clusters (Trentham; Driver \etal)
and flatter at the faint end than another 
($\alpha \approx -1.5$, Smith \etal),
(3) the composite group GLF rises more steeply at the faint end than the 
R-band GLF of the 
Las Campanas Redshift Survey
(LCRS; $\alpha = -0.7$ from Lin \etal), a large volume survey
dominated by galaxies in
environments more rarefied than luminous X-ray groups,
(4) the shape difference between the LCRS field and composite group
GLF's 
results mostly from the population of non-emission line galaxies
(EW [OII] $< 5$ \AA),
whose dwarf-to-giant
ratio is larger in 
the denser group environment than
in the field (cf. Ferguson \& Sandage, Bromley \etal), and
(5) the non-emission line dwarfs are more concentrated about the group
center than the non-emission line giants, except for the central, brightest 
($M_R < \Mstar$) group elliptical (BGG).  This last result indicates that the dwarfs,
giants, and BGG occupy different orbits ({\it i.e.}, have not mixed completely)
and suggests that some of the populations formed at a different times.

Our results show that the shape of the GLF varies with environment and
that this variation is due primarily to an increase in the
dwarf-to-giant ratio of quiescent galaxies in higher density regions,
at least up to the densities characteristic of X-ray luminous poor groups.
This behavior suggests that, in some environments,
dwarfs are more biased than giants with
respect to dark matter.  This trend conflicts with the prediction of
standard biased galaxy formation models.  

\bigskip
\bigskip
\noindent{\it Subject headings}:  galaxies: luminosity function --- 
galaxies: evolution --- galaxies: clusters: general --- 
cosmology: large-scale structure of Universe
}

\vfill\eject
\section{Introduction}
The shape of the galaxy luminosity function (GLF) in a given environment
is determined by the initial distribution of galaxy 
luminosities and by the subsequent galaxy luminosity and number density 
evolution.  Both the initial
luminosity and the luminosity/density evolution may depend on environment,
causing a variation in dwarf-to-giant ratio ($D/G$) with environment.
For example, the standard model of biased galaxy
formation predicts that giant galaxies are more likely than dwarfs to
form in regions of high mass density (cf. White \etal 1987).  
After galaxy formation, $D/G$ may be altered by mechanisms whose efficiency
is strongly environment-dependent,
{\it e.g.}, galaxy-galaxy mergers are probably more frequent and
global tidal fields weaker in poor groups than in rich clusters of galaxies
(cf. Zabludoff \& Mulchaey 1998, hereafter ZM98).  
A new class of cosmological models involving ``locally biased'' galaxy
formation (cf. Kauffmann \etal 1997;
Narayanan \etal 1998; Kravtsov \& Klypin 1999) 
has been introduced to modify standard biased galaxy formation
and to account for more complex environmental effects on
galaxy evolution.  Despite some recent
progress, observational uncertainties have prevented the behavior of
the GLF with environment from becoming a useful constraint.

Most observational determinations of the GLF to date have focused on
the field and rich clusters.  The GLF is even more uncertain in regions of
intermediate galaxy density, like poor groups, that are common galaxy
environments.  
To better constrain the models, and thus the relative
effects of environment-dependent galaxy formation and 
environment-driven galaxy evolution, we must ascertain (1) whether the
GLF of poor groups is universal, (2) whether the group GLF differs
from the GLF's of rich clusters and the field, (3) what galaxy
populations are most responsible for any environmental differences
({\it e.g.}, star forming or quiescent galaxies),
and (4) whether the GLF varies with local environment within a group
itself.

Past determinations of the shape of the poor group GLF differ widely.
Some composite group GLF's are consistent with the field GLF (Muriel
\etal 1998; Zepf \etal 1997), and others suggest a relative
depletion of faint galaxies (as reviewed by Hickson 1997) or a dip in galaxy
counts at $M_R \sim -18 + 5$log $h$
(Hunsberger \etal 1998).  Some
of the uncertainty arises because the number of known members per
group is often small: bound groups cannot be distinguished from chance
superpositions of galaxies along the line-of-sight, and the GLF cannot
be calculated without statistical background subtraction, a procedure
sensitive to inhomogeneities in the large-scale structure (especially for
low surface density contrast groups).
Furthermore, it is difficult to compare existing group GLF's with
those of the field and rich clusters, because previous
studies focus almost exclusively on Hickson Compact Groups (HCG's; Hickson 1982),
which are defined by their unusually concentrated bright galaxy
population and thus represent only one subset of groups in general.

The first step in addressing these problems is to identify a sample of
poor groups with (1) the properties of bound systems, {\it i.e.}, 
where there is
evidence that members lie in a common potential well, (2) a large
number of spectroscopically-confirmed members in each system, and (3)
galaxy environments different than those explored in past work.
In ZM98a and MZ98, we found that poor groups with luminous, extended X-ray
halos also have significant dwarf populations and that global
X-ray properties such as luminosity and temperature are well-correlated with
global optical properties like galaxy velocity dispersion.
These results argue that the members of an X-ray luminous group are
bound.

The large number of known members ($\sim 30$-60) in each X-ray group
not only renders background subtraction unnecessary, but also makes
statistically significant comparisons possible.  In particular, we can
learn whether this class of bound groups has a common GLF, and, if
not, what galaxy populations are responsible for the differences.
Furthermore, we can test whether the spatial distributions of distinct
galaxy populations 
{\it within} groups are consistent with any global, density-dependent
trends observed when comparing
GLF's of the field, poor groups, and rich clusters.

The shape of the GLF for members of poor, X-ray luminous groups also
provides insight into galaxy evolution in an environment that has not
been isolated previously.  Field studies such as the Las Campanas Redshift
Survey (LCRS; Lin \etal 1996) are dominated by galaxies in even more
rarefied environments than X-ray luminous groups ({\it i.e.}, by
members of poorer groups and by galaxies outside of associations).  
Many HCG's and other optically-selected poor group candidates 
do not have a hot, extended intragroup medium.  In contrast, some
properties of X-ray luminous groups are consistent with an
extrapolation of rich cluster properties to lower masses (MZ98;
ZM98b).  A direct comparison of the GLF's for these groups, rich
clusters, and the field has yet to be made.

In this paper, we combine multi-object spectroscopy and wide-field CCD
imaging of a sample of five nearby, X-ray luminous poor groups,
including three non-HCG's, to determine the form of the group GLF.  
For comparison, we also discuss the properties of a sixth
group, NGC 3557, that is marginally X-ray-detected.  We describe the
group sample, the photometry, and the spectroscopy in Section 2.
Section 3 contains the GLF determinations for individual groups, a
comparison of the composite GLF for the five X-ray luminous groups
with the GLF's for rich clusters and the field, an analysis of the
relative contributions of star forming and quiescent galaxies to the
differences between the group and field GLF's, and a comparison of the
spatial distributions of dwarf and giant group members.  Section 4
reviews some of the implications of our results for models of galaxy
formation and evolution.  Our conclusions are summarized in Section 5.

\vfill\eject
\section{The Observations}

\subsection{The Group Sample}
A poor group is defined optically as an apparent system of fewer than
five bright ($\simless \Mstar$) galaxies.  To isolate the form of the
GLF in poor groups with luminous X-ray halos, we examine five
X-ray-detected poor groups originally discussed in ZM98.
All five groups have extended ($>$ 100 h$^{-1}$ kpc), luminous
($L_X \sim$ 10$^{42}$ h$^{-2}$ erg s$^{-1}$) X-ray emission imaged by the 
ROSAT Position Sensitive Proportional Counter (PSPC) (Mulchaey
\& Zabludoff 1998; hereafter MZ98). 
For comparison, we obtain galaxy spectroscopy and
photometry for a sixth group, NGC 3557,
that is marginally-detected by ROSAT ($L_X = 2.8 \times$ 10$^{40}$ h$^{-2}$
erg s$^{-1}$)
and that has an asymmetric, unrelaxed X-ray morphology (Figure 1).
The X-ray temperature of NGC 3557 is also significantly lower 
($\sim$ 0.5 keV),
than is typical for the X-ray luminous groups ($\sim$ 1 keV),
although NGC 3557's temperature is poorly constrained due to the
group's relatively low X-ray luminosity.
Because NGC 3557 extends over an optical radius comparable to
that of the other groups, its lower temperature implies a lower
mass density.  This argument is supported by
NGC 3557's relatively low galaxy number density and velocity dispersion
(cf. Table 2).
The six groups have mean velocities of $2800 < cz <  7700$ \ks,
virial masses of $\sim 10^{13}-10^{14} \Mdot$, and a brightest group
galaxy (BGG) that is a giant elliptical located in the group center
(cf. ZM98a).

\subsection{Spectroscopic Data}
We obtained spectra for 742 galaxies in the six sample groups with the
multi-fiber spectrograph (Shectman \etal 1992) and 2D-Frutti detector
mounted on the du Pont 2.5m telescope at the Las Campanas Observatory.
Of these spectra, 328 are new observations, and the remainder are from
ZM98.  To define galaxy targets in each group field over the
$1.5\times 1.5$ degree$^2$ field of the fiber spectrograph, we used
coordinates, star/galaxy classifications, and relative magnitudes from
FOCAS (Jarvis \& Tyson 1981)
and the STScI Digitized Sky Survey.  The uncalibrated,
relative magnitudes drawn from the plate scans were sufficient to
identify the $\sim 200$ brightest galaxies in each field.
For each group, we observed 1-3 fiber fields, starting with the
brightest galaxies.
The completeness of the spectroscopic
sampling of each group field as a function of galaxy magnitude
is discussed in the next section and in $\S3.1$.

We determine radial velocities from the spectra using the
cross-correlation routine XCSAO and the emission line finding routine
EMSAO in the RVSAO package in IRAF (Mink \& Wyatt 1995).  The
velocities in Table 1 are either emission line velocities, absorption
line velocities, or a weighted average of the two (see Shectman \etal
1997 (their $\S2.2$) or Lin 1995 for a discussion of the
cross-correlation templates and the spectral lines typically
observed).  We compute velocity corrections to the heliocentric
reference frame with the IRAF/HELIO program.  See ZM98 for a dicussion
of the velocity zero-point correction and external velocity error
determinations.  

The distribution of galaxy velocites, the total number of galaxies
with velocities ($N_{tot}$), and the number of group members
($N_{grp}$) in each of the six fields are shown in Figure 2.

\subsection{Photometric Data}
We acquired images for the six groups under photometric conditions using 
the 40-inch telescope at Las Campanas Observatory during 
October 1996 and February 1997. 
The detector was a Tektronics 2048$^2$ CCD with a 
field of view of $\sim$ 23.8$^\prime$ on a side.
To cover the entire $1.5\times 1.5$ degree$^2$
area of our fiber spectroscopy field, we obtained a $5\times 5$ 
mosaic in all cases except for the more distant NGC 4325 group, for which
a $3\times 3$ mosaic 
was sufficient to image nearly all of the
spectroscopically-confirmed group members.
Each tile of the mosaic has
a $\sim 5^{\prime}$ overlap with an adjacent tile.

The total exposure time for each tile is 5 minutes with a Kron-Cousins
R filter from the Harris set. 
The typical seeing was $\sim$ 1.5$''$.  We reduce the
images using standard techniques in IRAF. The bias
level is determined from the overscan region of the CCD and 
subtracted from the images. Flat-fielding is accomplished 
using dome flats. The images are flux-calibrated using 
standard star fields in Graham (1982). 

Once the images are calibrated, we use the program 
SExtractor (Bertin \& Arnouts 1996) to classify objects as stars or galaxies
and to measure total magnitudes. For the purposes of this study,
we consider all objects with a ``stellarity-index" of less than
0.5 as galaxies. To verify that this classification is valid, we
examine plots of isophotal surface area versus magnitude for each field.
These plots indicate that the star/galaxy separation is typically valid 
down to $m_R \approx 19.5-20$. However, a 
small fraction of the images (less than 10\%) were taken under poor
seeing conditions ($\sim$ 2.5$''$). In these cases, the star/galaxy 
separation is less robust. To quantify the success of the SExtractor
classification for these fields, we visually classify the 
objects in one group, HCG 62. We find that the SExtractor classification
is consistent with our visual classification for all objects brighter than
$m_R = 18$. In the range $18 < m_R < 19$, the two methods 
yield consistent results 85\% of the time. 

In most cases, total magnitudes are measured using a method
similar to that proposed by Kron (1980). However, the 
Kron method relies on aperture magnitudes, which are
sensitive to crowding in the field. Thus, if a galaxy has nearby 
neighbors, the Kron magnitude may be inaccurate.
A better estimate of the true magnitude in these cases is a 
corrected isophotal magnitude (see discussion in the SExtractor manual).
Therefore, we adopt the `MAG\_BEST' option in
SExtractor, which computes a corrected isophotal magnitude when crowding
is a problem and a Kron magnitude otherwise. For the six group fields,
the Kron method is used to calculate the total magnitude in more
than 80\% of the galaxies.

We estimate the errors in the `MAG\_BEST' R magnitudes obtained
from SExtractor in several ways.  Because the CCD mosaic tiles
overlap, about 30\% of the galaxies are imaged more than once.  From
these multiple measurements, we estimate that the typical internal
magnitude errors are about 0.05 mag.  These errors are consistent with
the median of those output by SExtractor for galaxies brighter than
about $m_R = 17$. A few of the galaxies have previously measured total
magnitudes in the R band listed in the NED database.  A comparison of
these magnitudes with our data yields a median external error estimate
of about 0.15 mag. While total R magnitudes only exist for a handful
of our targets, many others have R-band aperature measurements in the
literature. A comparison of our photometry with that in the literature
in the same size aperture is consistent with our external error
estimate derived from the comparison of total magnitudes.

The completeness of the spectroscopic survey of each group field
is shown in Figure 3.
For each $m_R$ bin, we indicate 
the fractional completeness of the spectroscopic
data relative to the photometric catalog of SExtractor-identified galaxies.
In the case of NGC 4325 and of
NGC 5129, there are two distributions of $m_R$ --- one for
the entire spectroscopic/photometric catalog 
and the other sampled within a smaller radius of 0.6\lith\inv\ to make
it consistent with the sampling radii 
($\sim 0.4-0.6$\lith\inv\ Mpc) for the other groups.  Note that
we use the smaller radius sample for all subsequent analyses involving
NGC 4325 and NGC 5129.

As a complement to this paper, we have submitted a table of the
galaxies in each group field with measured velocities to the NASA/IPAC
Extra-galactic Database ((NED), Helou \etal 1991).  This table contains
the galaxy name, J2000 coordinates, heliocentric velocity and error,
type of velocity measurement ({\it i.e.}, from absorption lines ``0",
emission lines ``1", or a combination of both ``2"), and R-band total
magnitude for the 742 galaxies with measured velocities.
Table 1 shows an example of the format.  The full table is also
available in electronic form from the authors on request.

Table 2 summarizes the properties of the six sample groups,
listing the group name, projected centroid calculated from the coordinates
of the group members in J2000 (unweighted by galaxy luminosity), 
number of members ($N_{grp}$), mean
heliocentric velocity ($\overline \upsilon$), line-of-sight velocity
dispersion ($\sigma_r$), 
total X-ray luminosity ($L_X$), 
sampling radius for the photometry ($r_{samp}$; same as in Figure 4), number of
members within $r_{samp}$ (${N}^{\prime}_{grp}$), corrected number
density of galaxies with $M_R \leq -17 + 5$log $h$, within
0.4\lith\inv\ Mpc of the group center, and assuming spherical symmetry 
($n_{0.4}$), and dwarf-to-giant ratio for galaxies with $M_R
\leq -17 + 5$log $h$ and within 0.4\lith\inv\ Mpc ($D/G_{0.4}$; defined 
as in $\S3.1$).

\vfill\eject
\section{Results}

\subsection{Individual Group GLF's}
Is the GLF universal among poor groups of galaxies?  
For a sample consisting of five groups and the Virgo and Fornax clusters,  
Ferguson and Sandage (1991) argue that the 
early type dwarf-to-giant ratio
increases with the richness of the system.
However, as discussed by those authors,
the interpretation of their results is complicated by
the lack of spectroscopic data and inhomogeneities in the radial sampling
of the group and cluster images.
With spectroscopic surveys of galaxies in the fields of 
poor groups (ZM98a; MZ98; this paper), we can ascertain more directly
which groups are likely to be bound systems 
instead of chance superpositions and which galaxies are
group members instead of interlopers.  

Our earlier work 
suggests significant differences in $D/G$ as a function of local
mass density --- although the number of giant group members is
comparable, groups that are X-ray detected have 
higher velocity dispersions (200-450 \ks\ vs. $< 200$ \ks) and
larger memberships (20-50 galaxies vs. $< 10$ galaxies)
than non-X-ray-detected groups (also see
Hunsberger \etal 1998).  Unfortunately, the small
number of members in the non-X-ray groups prevents us from determining
if they are bound.  Therefore,
to test whether $D/G$ does vary with mass density,
we compare the individual group GLF's for a sample
of six X-ray-detected groups, including N3557, a marginal detection
and lower mass density environment.

The distribution of galaxy luminosities for each group is shown in Figure 4.  
The absolute magnitudes are calculated for a
$H_0 = 100$ \ks\ Mpc$^{-1}$, $q_0 = 0.5$ cosmology.  
For each group, we apply a global
extinction correction of $A_R = 0.58 A_B$, where $A_B$ is the
extinction in the B-band at the group's center
(NED) and the conversion
factor is estimated from the extinction curve of Schild (1977).
The GLF's are also corrected for incompleteness (see Figure 3)
by assuming that, within each magnitude bin, 
the fraction of galaxies without velocities that
are group members is the same as the fraction of measured galaxies that
are members.  Down to $M_R \leq -17 + 5$log $h$, the faint limit of our
subsequent analyses, the completeness corrections are small for each
group ({\it i.e.}, the corrected and uncorrected
counts are consistent within the $1\sigma$ counting errors). 
Note also that for these completeness corrections,
HCG 42, HCG 62, and NGC 3557 are $\simgreat 50\%$ 
complete within the $-17 < M_R \leq -16 + 5$log $h$ bin.

To test whether the distributions of galaxy luminosities differ among the
groups, we calculate a dwarf-to-giant ratio, $D/G$.  We define
giants as galaxies with $M_R \leq -19 + 5$log $h$ (corresponding to
$\simless \Mstar + 2.5$) and dwarfs by the range $-17 + 5$log $h \geq
M_R > -19 + 5$log $h$ (corresponding roughly to $\Mstar + 2.5$ to
$\Mstar + 4.5$, our faint end completeness limit; $\S3.2$)
\footnote{Note that unlike Ferguson \& Sandage (1991), who used galaxy
surface brightness to both assign group membership and to separate
early type dwarfs from giants, we separate the dwarf and giants
samples by galaxy luminosity.  Another difference is that the faint
end limits of our survey are typically $> 1.5$ magnitudes brighter.}.
To ensure that $D/G$ is calculated uniformly for all the groups, we
consider only members within 0.4\lith\inv\ Mpc of each projected group
center and completeness-correct the counts ($D/G_{0.4}$; see Table 2).

We calculate the errors in $D/G_{0.4}$ by assuming Gaussian counting
statistics (which are indistinguishable from the true Poisson errors
for all but the brightest bins in Figure 4).  The errors for the
completeness-corrected counts are determined with standard error
propagation.  The assumption of counting errors does not reflect an
intrinsic uncertainty in the number of galaxies in any magnitude bin
(for $M_R < -17 + 5$log $h$, the bins are complete or nearly so and
the only source of error is magnitude uncertainties).  Instead, the
errors provide estimates of how well the individual group GLF
determines the universal GLF (if it exists).  Because these errors
assume that all groups are drawn from the same parent GLF, they are
useful in testing whether the group $D/G_{0.4}$'s are statistically
different from one another.

The $D/G_{0.4}$ values for the five groups with X-ray luminous halos
are not statistically different.  However, the relative dearth of $-20
+ 5$log $h \geq M_R > -17 + 5$log $h$ galaxies in the galaxy
luminosity distribution of NGC 3557 compared with the X-ray luminous
groups produces a lower $D/G_{0.4}$.  The composite $D/G_{0.4}$ of the
five X-ray luminous groups (computed by normalizing each group's GLF
to that of HCG 42 and averaging; see $\S3.2$) is $1.9 \pm 0.4$, which
differs at the $> 4\sigma$ level from NCG 3557's value of $0.2 \pm
0.2$.  For comparison, the Local Group's $D/G$ is roughly $< 0.8$ in
this magnitude range (Grebel 1999).

Although $D/G_{0.4}$ is lower in the NGC 3557 group than in 
the other groups down to our completeness
limit of $M_R < -17 + 5$log $h$, NGC 3557's
galaxy luminosity distribution rises at fainter magnitudes 
(even the uncorrected, lower-limit counts rise).  Deeper spectroscopic
surveys of the other, more distant groups will determine
whether the behavior of their extreme faint end GLF's is similar to that
of NGC 3557.  The ``dip" in
NGC 3557's GLF is roughly consistent in shape with the composite
GLF for mostly non-X-ray luminous Hickson Compact Groups observed by
Hunsberger \etal 1998, who suggest that dynamical friction
and galaxy mergers cause intermediate
luminosity galaxies to acquire mass and
to move to the bright end of the GLF in some poor groups.  The low specific
globular cluster frequency and high rotational velocity of 
NGC 3557 itself are consistent with
a merger product (van den Bergh 1986).  
Additional explanations
for differences among the GLF's of groups are discussed in
$\S4$.

The results of this section suggest that the
GLF is not universal among poor groups and that
$D/G$ may increase with the mass density of the group environment.
In the next section, we examine whether this trend in $D/G$ continues
from the field to poor groups to rich clusters.

\subsection{Composite Group GLF}
Because luminous, extended X-ray emission suggests a common
potential well and is roughly correlated with the number density of
group galaxies, the fraction of early type members, 
and the group velocity dispersion (ZM98; MZ98), a sample of
X-ray luminous groups is likely to contain a higher fraction of
bound systems than a sample of
group candidates identified only as galaxy concentrations in velocity space
and on the sky.
As a result, determining 
the GLF for X-ray luminous groups is an important first
step in isolating the effects of group environment on the evolution of
galaxies.  With our sample, it is
now possible to compare the shape of the GLF in three 
{\it distinct} environments: the Las Campanas Redshift Survey
of the field, X-ray luminous groups, and rich clusters of galaxies.

We construct a composite GLF 
from the five groups with luminous X-ray halos
by arbitrarily normalizing each to have 
the same number of completeness-corrected galaxy counts
brighter than $M_R = -17 + 5$log $h$ as HCG 42.  
We then
average the five individual, completeness-corrected group GLF's.  This procedure ensures 
that the shape of the composite GLF is not weighted more by groups
like HCG 62 and NGC 2563, which have relatively high
galaxy densities.  Because HCG 42 and HCG 62 are
complete to within a factor of two for galaxies with
$-17 +5$log $h \geq M_R \geq -16 + 5$log $h$, we average the corrected galaxy counts
for these two groups only to obtain the composite point for that bin.
However, only the five bins
brighter than $M_R = -17 + 5$log $h$ are used in subsequent determinations of $D/G$.

The composite GLF for group members with 
$-23 + 5$log $h < M_R < -16 + 5$log $h$ 
and within projected radii of $\simless 0.4-0.6$\lith\inv\ Mpc
from the group center 
is consistent with a Schechter function of form
$\Mstar = -21.6 \pm 0.4 + 5$log $h$ and $\alpha = -1.3 \pm 0.1$ (Figure 5).  
Figure 5 also shows two composite GLF's for
rich clusters (Trentham 1997 (includes Coma),
Driver \etal 1998) and the GLF for the Las Campanas
Redshift Survey (hereafter LCRS) of the field (Lin \etal 1996).
The field and cluster GLF's are 
normalized so that the number of galaxies
brighter than $M_R < -17 + 5$log $h$,
roughly the completeness limit for the group and
LCRS samples, is the same as for the group composite.
The completeness of the cluster samples, as derived from 
background-subtracted, not spectroscopic, counts,
is estimated to be one or two magnitudes fainter.
The two cluster GLF's, which are calculated from different cluster samples,
are consistent with one another within the errors except in
the $M_R = -21.5 + 5$log $h$ bin.  The cluster galaxies in this
bin contribute little ($\sim 2\%$ for Trentham, $\sim 12\%$ for Driver
\etal) to the total number of giants with $M_R < -19 + 5$log $h$, and
thus the D/G's of the two cluster samples are similar.

Although we have measured most or all of the group members brighter
than $M_R < -17 + 5$log $h$ in each group,
the statistics of the group sample 
are not adequate to distinguish among
different functional forms for the composite 
GLF ({\it e.g.}, between the single Schechter
function and a two-component fit (cf. Hunsberger \etal 1998).  
Over this magnitude range,
the cluster and LCRS GLF's are also well fit
by single Schechter functions, and 
the composite group data, not the fit, are compared to
these GLF's in the next two sections.  Even the addition of NGC 3557, whose
GLF suggests a non-Schechter functional form (cf. Figure 4), 
does not significantly alter the composite ({\it i.e.,} the Schechter
function is not excluded by a $\chi^2$ fit, and $\Mstar$
and $\alpha$ are within the original $1\sigma$ errors).

We stress that the field, group, and cluster GLF's in Figure 5
do not represent the absolute contribution of each environment to
a unified GLF, as the normalizations are arbitrary.  Instead,
the shape of each GLF suggests the typical luminosity distribution of member
galaxies in that environment.  To properly normalize the composite group GLF
for bound groups over the mass range of our sample
($\sim 10^{13}$-$10^{14} \Mdot$),
we would need to know what fraction of groups cataloged from
optical redshift surveys are bound and what fraction of bound groups
have luminous X-ray halos, marginal X-ray detections, or non-X-ray-detections.

\subsubsection{Comparison with Rich Clusters}
How do the GLF's for poor groups compare with
those for rich clusters of galaxies?  Determinations of
individual cluster GLF's
vary in part due to differences in observed
waveband (cf. Wilson \etal 1997) and in mean sample redshift, factors that are
sensitive to morphology and/or star formation history.
However, as in the case of poor groups,
there is evidence for intrinsic
variations among the GLF's of rich clusters (cf. L\'opez-Cruz \etal 1997;
Driver \etal 1998).
To investigate whether there are global trends in the shape of the GLF
from poor groups to rich clusters, we compare our R-band composite GLF
for X-ray luminous groups with three composites of nearby ($z \leq 0.2$)
clusters in the R-band (Trentham 1997, Smith \etal 1997, and Driver \etal 1998).

Figure 5a shows that our composite GLF and those
derived from four rich clusters 
(Trentham 1997) and from seven other rich clusters (Driver \etal 1998)
are consistent down to 
$M_R < -16 + 5$log $h$ and also over the extrapolation of the group GLF
one magnitude fainter (a $\chi^2$ test is unable to distinguish
at the $>95\%$ confidence level
among the three GLF's over these magnitude ranges).
Smith \etal derive a somewhat steeper faint
end slope from a composite of three rich clusters
($\alpha \approx -1.5$ vs. $\alpha \approx -1.3$).
The Trentham and Driver \etal GLF's are constructed from Kron
total magnitudes (Smith \etal use isophotal magnitudes), which are
typically equivalent to the SExtractor
`MAG\_BEST' magnitudes ($\S2.3$) in our group GLF.
All three cluster GLF's are determined using
statistical background subtraction.

In fitting their cluster composite,
Smith \etal obtain a slightly better fit using
two Schechter functions instead of one (both the one- and two-Schechter 
fits have steeper faint end slopes than Trentham and
Driver \etal over the magnitude range
of our data).
Smith \etal include the Coma cluster, which has either a sharp rise
that exceeds a single Schechter function at faint magnitudes or an actual dip
due to a relative deficit of dwarf galaxies with
$M_R \sim -18 + 5$log $h$ (Secker \& Harris 1996).  
The Coma cluster is highly complex, with several recently
accreted groups.   If some of these infalling groups
have GLF's that are better fit by two components (cf. Hunsberger \etal 1998;
Koranyi \etal 1998), the overall shape of the Coma GLF may be determined
in large part by the
contributions of those subclusters (cf. Secker \& Harris 1996).
This suggestion is supported by the consistent
dwarf-to-giant ratios of Coma and of the similarly complex, but
poorer, Virgo cluster (Thompson \& Gregory 1993).

Differences among the GLF's of individual groups and rich clusters
might result from an environment-dependent 
combination of type specific GLF's
(Binggeli, Sandage, \& Tammann 1988; Jerjen \& Tammann 1997).  
It is also possible that
two galaxies of the same initial morphology might experience different density
and/or luminosity evolution depending on their environment,
leading to evolution in the type-specific GLF's.
For example, L\'opez-Cruz \etal (1997) and Driver \etal (1998) 
find a different trend among clusters than we do among groups and
than Thompson \& Gregory (1993) and Valotto \etal (1997) 
find among other clusters ---
namely, that dwarf-to-giant ratio decreases with increasing global projected
galaxy density.  Driver \etal observe the effect 
only outside the core,
a region more sensitive to background subtraction and cluster substructure.
However, their result suggests that
the trends in $D/G$ among poor groups and
from the field to poor groups may be reversed
in some rich clusters by a different galaxy evolution history.

While the details of the inter-dependence of galaxy type, luminosity,
and environment await future surveys,
we conclude here that the typical
$D/G$ of rich clusters is either consistent with (cf. Trentham 1997,
Driver \etal 1998) or
larger than (cf. Smith \etal 1997) that of X-ray luminous poor groups.

\subsubsection{Comparison with LCRS Field}
Estimates of the luminosity function of galaxies in the nearby ($z
\sim 0.1$) field vary as much as the observed GLF's for rich cluster
members.  As in the case of the cluster GLF, the uncertainty in the
field GLF arises in part from the difficulty in translating the
different photometric filters employed by redshift surveys into the
same band.  Such translations may ignore potentially important effects,
including the initial selection of galaxies from different bands,
variations in galaxy color with absolute magnitude ({\it e.g.}, the
``mass-metallicity relation''), and intrinsic differences between the
dwarf-to-giant ratios of blue and red galaxies (witness the
differences between the emission and non-emission line GLF's discussed
in $\S3.3$).  It is therefore essential to compare our R-band
composite group GLF with a R-band GLF of the field.

Another issue is how fairly a given survey samples the nearby
universe.  For example, it is possible that high density environments
are overrepresented in the R-band CfA Century survey (Geller \etal
1997), which contains portions of the Corona Borealis supercluster and
of seven Abell clusters (including Coma).  In contrast, the larger,
R-band Las Campanas Redshift Survey (LCRS; Shectman \etal 1996)
is known to be dominated by galaxies in environments more 
rarefied than X-ray luminous groups.
Fully $87\%$ (18590 out of 21343; cf. Tucker \etal 1998) of LCRS galaxies 
lie outside of poor groups or in groups
that have lower velocity dispersions ($\leq 200$ \ks),
and presumably lower mass densities, than groups in our X-ray
sample.  Although this fraction may be overestimated relative to the
``true" field due to that fiber survey's
tendency to undersample overdense regions,
the LCRS is an appropriate choice for comparing
our GLF for X-ray luminous, poor group members
with that for galaxies in typically less dense environments.

Figure 5b shows the GLF of the
composite of the X-ray luminous groups and the
best Schechter fit to the LCRS field survey GLF (Lin \etal 1996).
If the arbitrary LCRS normalization is adjusted to
minimize $\chi^2$ with respect to the group GLF for galaxies
brighter than the estimated LCRS completeness limit of 
$M_R = -17.5 + 5$log$h$ (Lin \etal 1996), the LCRS field GLF is excluded at the
$>95\%$ level.  (The best $\chi^2$
normalization is in fact lower than that shown).
Relative to the field, poor groups with luminous X-ray halos have 
either a deficit of giants, an excess of dwarfs,
or a combination of both effects.

The difference between the LCRS field and the poor group composite
is not due to the difference between the isophotal magnitudes used
in the LCRS GLF and SExtractor `MAG\_BEST' magnitudes calculated for
the group members.  From the galaxies in our sample with
$10.3 \leq m_R \leq 17.3$ (the magnitude range used to calculate the
group GLF),
we estimate that the isophotal to `MAG\_BEST' magnitude correction
to the LCRS GLF is typically $\simless -0.2$.
This value is consistent with
that estimated by Lin \etal ($-0.35 \pm 0.1$; 1996).
Applying this correction, which increases slightly towards fainter magnitudes,
only furthers the disagreement between the LCRS field and the composite group
GLF's in Figure 5b.

Incompleteness in the LCRS is unlikely to be the source of the trend
towards higher $D/G$ in the denser, group environment.  First, we
compare the group and LCRS samples only down to the estimated $M_R$
limit above which the LCRS is completeness-corrected 
(Lin \etal 1996)\footnote{We address the possibility of type-dependent
incompleteness in $\S3.3$.}.
Second, the observed increase in $D/G$
with density is consistent with the results of an analysis of
the LCRS itself (Bromley \etal 1998), 
where any faint incompleteness in the galaxies
would be either uniform across the sample or greater in higher
density regions.  
Third, it is suggestive
that the only other large R-band survey of the nearby
field (the CfA Century survey, Geller \etal 1997) has both a
higher average galaxy density and a larger dwarf-to-giant ratio than the LCRS
($\alpha = -1.2$ vs. $\alpha = -0.7$, respectively).  

In summary, $D/G$ increases from the LCRS field, which is dominated by
galaxies in poorer groups and outside of groups, to groups with
X-ray luminous halos.

\subsection{Star Forming vs. Quiescent GLF's}
Is it possible to isolate the galaxy population most responsible for the
increase in the $D/G$ between the field and X-ray luminous groups?
Ferguson \& Sandage (1991) suggest
that the differences between the dwarf-to-giant ratios
of groups and clusters are due mostly to an
increase in the early-type dwarf-to-giant
ratio with richness.  A recent analysis of the LCRS (Bromley \etal 1998) using
spectroscopically-defined galaxy morphologies also finds that the
early type dwarf-to-giant ratio increases with local density.
By analyzing the emission line characteristics
of the group and LCRS galaxies, we can divide the data into star forming
and non-star forming (quiescent) galaxies.  As in Lin \etal 1996, we define
star forming group members as those with [OII] EW $> 5$\AA\ (approximately
the Galactic value).
Galaxies with a weaker or non-detectable [OII] line are classified
as quiescent.  

The GLF's for the divided samples are shown in Figure 6.
For both the LCRS and the group samples, the GLF for star forming galaxies
rises more steeply than that for quiescent galaxies.  
The two GLF's for the LCRS 
sample are each normalized to have the same number of galaxies
brighter than $M_R = -17 + 5$log $h$ 
as the corresponding composite group GLF's.

For the five brightest bins (corresponding roughly to the 
LCRS completeness limit), the quiescent galaxies
in the LCRS field and in the X-ray
luminous groups have different GLF's, {\it i.e.,}
adjusting the relative normalizations
to minimize $\chi^2$ excludes the field sample
at $> 95$\% confidence.
The $\chi^2$ minimization also forces the
normalization lower than plotted,
increasing the differences between the 
field and the groups at the faint end.
In contrast, the star forming galaxies
have roughly consistent GLF's down to the 
$M_R \sim -17.5 + 5$log $h$ bin (the $\chi^2$ minimization test 
does not distinguish between the
two star forming GLF's).  

One potential problem in interpreting these results
is that noise in a spectrum
can be mistaken for an [OII] emission line.
Therefore, in the case of
low signal-to-noise spectra ({\it i.e.}, dwarfs), it is possible to
overestimate the number of star forming galaxies.
We test the magnitude of this effect by applying an [OII] flux cut
of $> 2\sigma$ to the star forming sample.  Although the number of 
group members classified as star forming is reduced from 48 to 26,
and the number of quiescent galaxies is correspondingly increased,
the resulting GLF's are consistent with those  
in Figure 6.

Another consideration is that the mean redshift of the LCRS galaxies is
higher than for the group sample ($z \sim 0.1$ vs. 0.017,
respectively).  As a result, the fixed $3.5^{\prime\prime}$ size of the
spectroscopic fiber subtends, on average, different physical radii for
the LCRS and group samples.  Because of this aperture bias,
light is sampled within the inner $\sim 1$\lith\inv\ kpc of a group galaxy
at the average survey depth,
in contrast to the $\sim 4$\lith\inv\ kpc 
sampling typical of LCRS galaxies.  However,
aperture bias is unlikely to significantly affect the star forming/quiescent
galaxy classifications and the disagreement between the group and field
GLF's for the following reasons.  The dominant effect of
aperture bias would be to prevent the detection of
HII regions in the disks of group members,
causing some star forming galaxies to be misclassified as quiescent.
This problem is rare because the effect is
only significant for face-on galaxies 
(inclined disks tend to have HII regions
along the line-of-sight).  For example,
few emission line spirals are classified as non-emission line galaxies
(about 1 of 12 within 15000 \ks; Zaritsky, Zabludoff, and Willick 1995).
Not only are the effects of aperture
bias on the GLF's (arbitrary) normalization small, but they are
unlikely to alter the GLF's {\it shape}, the basis of our comparison of the
group and field populations.

One way to artificially reproduce the trends in Figure 6 is to
stipulate that many faint emission line dwarfs are missing from the
LCRS and that environmental conditions in groups convert them to
non-emission line dwarfs.  However, this model is problematic.
First, such a transformation between star forming and quiescent
dwarfs is unlikely.  Although
mechanisms like tidal stripping or ``galaxy harassment" (Moore \etal
1996; Moore \etal 1998) have been proposed for transforming star
forming irregulars or Sd's into quiescent spheroidals, studies of
dIrr's and dE's in Virgo show that the structures defined by the old
stellar populations differ significantly between the two types of
dwarfs.  For example, the dIrr's have more flattened and asymmetric
stellar light distributions, and no dIrr's have H-band luminosities or
surface brightnesses as high as those of the brightest dE's (James
1991).

Second, incompleteness in the LCRS does not affect our results significantly.
Huchra (1999) argues from the
B-band CfA2 redshift survey that the LCRS selection criteria exclude
more faint, low surface brightness galaxies than are corrected for by
Lin \etal (1996) and that these galaxies have mostly emission line
spectra.  Even if it were simple to compare B-band data directly with
the R-band LCRS (and, for the reasons cited earlier, it is not), the
following argument suggests that the effects of any missing galaxies
are small by showing that the combination
of incompleteness and of dwarf transformation leads to consequences that we do
not observe.

Is it possible to transform the field GLF into the group GLF by
changing field emission line dwarfs into group non-emission line
dwarfs?  We define the total number of field galaxies at $M_R = -17.5
+ 5$log $h$ that will become group galaxies as $E_i + N_i$, where
$E_i$ and $N_i$ are the number of star forming and quiescent dwarfs,
respectively, in the field.  The final number of group galaxies in the
$M_R = -17.5 + 5$log $h$ bin is then $E_f + N_f = E_i + N_i$, where $E_f$
and $N_f$ are the star forming and quiescent group dwarfs,
respectively.  First, we correct for the ``missing" field dwarfs.  The
difference between the CfA2 and LCRS emission line galaxy counts at
$M_R = -17.5 + 5$log $h$ is a factor of $\sim 4$ (Huchra 1999).  The
difference between the LCRS emission line and non-emission line counts
in this bin is a factor of $\sim 6$ (Lin \etal 1996, as opposed to the
arbitrary relative normalization shown in Figure 6).  If we
``correct'' the LCRS emission line counts by the CfA2 value, the ratio
of emission to non-emission line counts in the field is $\sim 24$ at
$M_R = -17.5 + 5$log $h$, {\it i.e.}, $E_i = 24 N_i$.  Second, we
measure the ratio of group emission to non-emission line dwarfs.
Figure 6 shows that $E_f \sim 1/2 N_f$.  Therefore, $3/2 N_f = 25
N_i$, and this model predicts that the final ratio of quiescent dwarfs
in groups to those in the field would be $\sim 17$ at $M_R = -17.5 + 5$log
$h$, while the ratio for giants does not change.  This result is at
odds with Figure 6, in which only a boost of at most $\sim 5\times$ in
the number of quiescent field dwarfs relative to giants is required to
match the observed group population.  The model, in which many faint
emission line dwarfs are missing from the LCRS and are converted
by group environment into non-emission line dwarfs, 
over-predicts the ratio of quiescent group dwarfs to quiescent group giants.
Because the model is wrong,
the effects of LCRS incompleteness on our results are likely
to be small.

In summary, we find in this section
that the quiescent $D/G$ in groups is significantly
larger than that of the field.  This result indicates that quiescent
dwarfs are more clustered than quiescent giants, although it is not
clear whether an excess of dwarfs, a deficit of giants, or some combination
of both effects, is responsible.

\subsection{Spatial Distribution of Dwarfs vs. Giants}
In previous sections, we find that $D/G$ increases with mass density
among groups, that $D/G$ increases from the field to groups
(and may continue to increase from groups to clusters), and that a
change in the $D/G$ of non-star forming galaxies is the cause
of the increase from the field to groups.  Therefore, if this trend
is real, we might expect $D/G$ to increase {\it within} groups from
the outskirts to the denser core.  Such behavior
would be opposite to the effect of mass segregation
and to the prediction of standard biased galaxy formation.
Ferguson \& Sandage (1991) identify no radial gradients
in the surface brightness-defined dwarf-to-giant ratio
in their study of the Virgo, Fornax, and Antlia systems. 
The luminosity-defined 
dwarf-to-giant ratio of rich clusters in the Driver \etal (1998) sample
rises from the inner ($r \leq 0.28$\lith\inv\ Mpc)
to outer ($0.28 < r \leq 0.37$\lith\inv\ Mpc)
annulus for some systems and falls for others.
Here we test for such gradients in our luminosity-defined $D/G$.

First we compare the kinematic and spatial distributions of the
BGGs, dwarfs, and giants for all six groups
(Figure 7ab).  Figure 7a is the
composite phase space diagram for the 123 quiescent galaxies.
The y-axis shows the velocity offset of the galaxy
from the mean velocity of the group, the x-axis shows the projected
radial offset from the projected group centroid normalized by
the group velocity dispersion.  The six sample
galaxies with $M_R < \Mstar$.
are marked by asterisks and include four BGG's\footnote{The 
BGG of NGC 4325 is star forming, and the
BGG of HCG 62 is fainter than $\Mstar$.}, which are consistent
with the kinematic and spatial center of their groups (ZM98a).
There is also an apparent concentration of
dwarfs (small filled circles) toward the group center relative to
giants (large open circles).

To compare the distributions of the samples on the sky and in
velocity space simultaneously and quantitatively, we define the statistic 
$R^2 = (x/\delta_x)^2 + (|y|/\delta_{|y|})^2$, where
$\delta_x$ and $\delta_{|y|}$ are the $rms$ deviations in $x$ and $|y|$
for the entire sample (cf. ZM98).
Thus, a galaxy that has a large peculiar motion
and/or that lies outside the projected group core will have a larger $R$ 
value than a galaxy at rest in the center of the group potential.
The distributions of $R$ for the four
BGGs and two other $M_R < \Mstar$
galaxies (heavily shaded), 56 giants (shaded), and 61 dwarfs (unshaded)
are in the right-hand panel.  A Kolmogorov-Smirnov test 
indicates that the dwarf and giant
(and the $M_R < \Mstar$ and giant) samples differ from one another at
the $>95\%$ confidence level.
No one group is responsible for
this difference ({\it e.g.}, removing NGC 3557, the marginally
X-ray-detected group, does not affect the 
outcome).  This result suggests that the BGG, dwarf, and giant
populations occupy different
orbits ({\it i.e.}, have not mixed completely).  

Figure 7b suggests that
the 49 galaxies with significant [OII] emission tend to lie outside
the group core and to have larger peculiar velocities than the
quiescent galaxies.  In fact, the overall $R$ distribution in Figure 7b
differs from that in Figure 7a at the $>95$\% level. 
As in the case of the
quiescent galaxies, the $R$ distributions for the 36 emission line dwarfs
and 12 emission line giants are significantly
different (at the $> 95\%$ level).
The $R$ values for the dwarfs are typically
smaller (also as in Figure 7a), implying that 
the star forming dwarfs are more concentrated radially and/or in velocity
space than the star forming giants.

To examine how $D/G$ varies with radius, and thus with mass density,
we focus on the larger sample of quiescent galaxies.  Figure 8 shows
$D/G$ in three radial bins for the quiescent galaxies of each group
in Figure 7a.  A Spearman rank-order test yields a strong correlation
coefficient of $-0.62$, which is significant at the $> 95\%$ level.
(The middle point for HCG 42 is not plotted, because the group has no
giant members within this annulus.  However, if we assume conservatively
that the missing point has the highest rank $D/G$ in the sample,
the Spearman coefficient is still significant at the $> 95\%$ level.)
The trend is likely to be even steeper than shown in Figure 8, because
the sample includes two Hickson Compact Groups, 
which have unusually low core $D/G$ values 
(the two lowest filled circles in the first bin) due to 
Hickson's (1982) selection criteria.
Removing the marginally X-ray detected group NGC 3557 (open circles), 
which is sampled only to 0.4\lith\inv\ Mpc and has the lowest $D/G_{0.4}$,
increases the steepness of the trend and the significance of the 
Spearman correlation coefficient.

The results of this section show that the dwarf and giant populations are
not well-mixed and that $D/G$ decreases with radius, and therefore
increases with mass density, within the group environment.  Mass
segregation, in which bright galaxies are brought via
dynamical friction into the group core,
would produce the opposite trend.  However, mass
segregation might
lead to mergers with the BGG that would disguise its effects.
While these results do not include evidence for
mass segregation, there are implications for models of standard biased
galaxy formation that we discuss in the next section.

\section{Discussion}
Our results suggest that dwarf-to-giant ratio increases with
the mass density of the environment.  This trend exists 
among poor groups, from the field to groups and
rich clusters (at least up to the densities of X-ray luminous poor groups), 
and within the groups themselves.
How might we explain the dependence of $D/G$ on environment, an
effect that runs counter to the prediction of standard biased
galaxy formation?
Empirically, we know that there is some relationship between a galaxy's morphology
and the density of its environment (Dressler 1980).  
It is also observed that the surface density of dwarfs projected within
$\sim 250$\lith\inv\ kpc of giant ellipticals is at least $3\times$
that around giant spirals
(Lorrimer \etal 1994).  Therefore, the combination of these two effects alone
would lead us to expect a boost in $D/G$ with environmental
density.  

While a morphology-density relation may be a natural consequence of
standard biased galaxy formation ({\it i.e.}, the most massive
galaxies, giant ellipticals, form preferentially in the dense
environments of clusters), the relative excess of dwarfs
around giant ellipticals is not.  The latter effect may instead be due
to an environmental variation in the efficiency of galaxy formation or
in the frequency of galaxy-galaxy mergers.  
To date, there are few
detailed theoretical models of such environmental/morphological
influences on $D/G$.  Scenarios that increase $D/G$ include:  1) giant
galaxies form less efficiently in denser environments (cf.
David \& Blumenthal 1992), and
dwarfs are the leftover material, 2) cold HI clumps ({\it
e.g.}, the High Velocity Clouds in the Local Group (Blitz \etal 1998))
are more likely to collide, produce stars, and evolve into dwarfs in
denser regions, 3) galaxy mergers, which occur 
more frequently in dense systems, reduce the giant population and
transfer both progenitors' satellites to a single remnant,
4) galaxy mergers produce tidal tails in which additional dwarfs form
(Barnes \& Hernquist 1992; Hunsberger \etal 1996), and
5) dynamical friction in
denser environments increases the merger rate of giants with
the central, giant
elliptical, which then acquires their satellites.
Although it is not
possible to distinguish among these possibilities at present, we note
that the non-mixing of the BGG, dwarf, and giant populations, in
addition to the clustering of dwarfs about the central BGG, suggests
that at least one of these populations evolved later than the others.

There is preliminary evidence that $D/G$ has evolved in
other nearby environments.  For example, in the simple environments
of isolated, giant elliptical galaxies
(cf. Mulchaey \& Zabludoff 1999; Colbert, Mulchaey, \& Zabludoff 1999),
we have found indications of mergers.  The
giant elliptical NGC 1132 has a poor group-like 
X-ray halo and dwarf population,
yet there are no other giant galaxies in its field.  This result
is consistent with the picture that NGC 1132 is a merged group.

The consistency of the dwarf-to-giant ratio for the clusters in
Trentham's (1997) and Driver \etal 's (1998) samples with that of the
X-ray luminous groups is reminiscent of another surprise in the
comparison of groups and rich clusters.  Zabludoff \& Mulchaey (1998)
find that some X-ray groups have early type galaxy fractions similar
to those of clusters, despite the lower velocity dispersions of the
groups.  The strong correlation between velocity dispersion and early
type fraction in groups thus deviates from linearity at cluster
velocity dispersions.  This saturation point occurs at a velocity
dispersion of 400-500 \ks, the value that a
poor group would require to enable an $M^*$ galaxy member to
experience a merger within a Hubble time.  Therefore, it is possible
that mergers cause some evolution in the early type fraction of poor
groups and cease to be effective in richer groups and clusters.  The
apparent saturation of dwarf-to-giant ratio with system density
observed here may be a manifestation of the same phenomenon.

The results of this paper are inconsistent with 
the prediction of standard biased galaxy formation models,
in which galaxy formation is modulated coherently over scales larger than
the galaxy correlation length,
and further motivate ``local biasing" models (cf. Narayanan \etal 1998),
in which the efficiency of galaxy formation is determined 
by the density, geometry, or velocity dispersion of the 
local mass distribution.

\section{Conclusions}

We use multi-object spectroscopy and wide-field CCD imaging to examine
the shape of the galaxy luminosity function (GLF) in six poor groups
of galaxies.  Five of these groups have luminous X-ray halos and thus
represent an environment in which the GLF has never been isolated.  
For these five groups, the composite group GLF for galaxies with $-23 +
5$log $h < M_R < -16 + 5$log $h$ and within projected radii of
$\simless 0.4-0.6$\lith\inv\ Mpc from the group center is consistent with a
Schechter function with $\Mstar = -21.6 \pm 0.4 + 5$log $h$ and
$\alpha = -1.3 \pm 0.1$.

Our other conclusions are:

1.  {\it The GLF is not universal in poor groups}.  The ratio of
dwarfs ($-17 + 5$log $h \geq M_R > -19 + 5$log $h$) to 
giants ($M_R \leq -19 + 5$log $h$) 
is significantly larger for the five luminous X-ray
groups than for the one marginally X-ray detected group.
The difference between the X-ray properties of NGC 3557
and the X-ray luminous groups may reflect a difference in their
potential well depths, as only deep wells heat gas
to X-ray-detectable levels (cf. ZM98, MZ98).
Because all of the groups have roughly the same
physical scale, this result suggests that
$D/G$ increases with mass density for these systems.

2.  {\it The dwarf-to-giant ratios of X-ray luminous groups are
consistent with or smaller than those for rich clusters}.  The
composite GLF for the luminous X-ray groups is consistent in shape
over the full magnitude range with two measures of the composite GLF
for rich clusters (Trentham 1997; Driver \etal 1998) 
and flatter at the faint end than
another ($\alpha \approx -1.5$, Smith \etal 1997).  This result
suggests that {\it if} there is any shape difference between the poor
group and rich cluster GLF's, it arises from a larger dwarf-to-giant
ratio in the denser cluster environment.

3.  {\it Dwarf-to-giant ratios are larger in X-ray luminous groups
than in regions outside of groups and in poorer groups}.  The shapes
of our composite group GLF and the large volume, R-band, Las Campanas Redshift Survey
field GLF (Lin \etal 1996) differ at the $> 95\%$ level.  The
shape difference is due either to an excess of dwarfs, a deficiency of
giants, or a combination of both effects in poor X-ray groups.
Because the LCRS is dominated by galaxies in environments more
rarefied than those of these groups, this result suggests that
$D/G$ increases with mass density from the field to
X-ray luminous groups.

4.  {\it Quiescent galaxies cause most of the difference
between the dwarf-to-giant ratios of X-ray luminous groups and the field}.
The GLF for emission line galaxies
(EW [OII] $> 5$ \AA)
in the X-ray groups 
is indistinguishable from that of the LCRS field.  On the other
hand, the GLF's for non-emission line galaxies in the groups and
in the field differ at the $> 95\%$ level.
Thus, the shape difference between the overall field and group GLF's (and
presumably between the field and rich cluster GLF's) is due mostly to the
population of quiescent galaxies, 
whose $D/G$ is larger in the denser
group environment than in the field (cf. Ferguson \& Sandage 1991, Bromley
\etal 1998).  

5.  {\it Quiescent dwarfs are more concentrated about the group center
than quiescent giants, except for the central, brightest
($M_R < \Mstar$) elliptical}.  A comparison of the velocities and projected
positions of the brightest group galaxies (BGG's), giants, and 
dwarfs in the X-ray groups suggests that
these populations occupy different orbits ({\it i.e.}, have not mixed completely)
and may have evolved via different mechanisms and at different times.
Furthermore, the group $D/G$ decreases with radius and therefore
increases with mass density.  

Our results show that the shape of the GLF varies with environment and
that this variation is due primarily to an increase in the
dwarf-to-giant ratio of quiescent galaxies in higher density regions,
at least up to the densities characteristic of X-ray luminous poor groups.
This behavior suggests that, at least in some environments,
dwarfs are more biased than giants with
respect to dark matter.  This trend is in conflict with the prediction of
standard biased galaxy formation models.  
If more than standard biased formation
is at work, then possible explanations include inefficient galaxy formation
({\it e.g.}, giants form less efficiently in denser environments),
increases in the satellite-to-primary ratio through the mergers
of giant galaxies, and dwarf formation in the tidal tails of giant merger remnants
(cf. Hunsberger \etal 1996).

\vskip 0.3in\noindent 
We thank the referee, John Huchra, for his careful reading of the manuscript.
We thank Dennis Zaritsky for his comments on the text and for helpful suggestions.
We also thank Huan Lin and Michael Vogeley for providing software
used in some of our analyses, Neil Trentham and Huan Lin for providing
electronic copies of their data tables and for useful discussions, 
and Romeel Dav\'e, Simon Driver, Neal Katz, Joel Primack, Ian Smail, David Spergel, 
and Scott Trager for important information.  This paper is based
on observations made at Las Campanas Observatory, Chile.
AIZ acknowledges support from NSF grant AST-95-29259 and NASA grant
HF-01087.01-96A.
JM acknowledges support provided by NASA grant NAG 5-2831 and NAG 5-3529.

\vfill\eject
\begin{figure}
\plotone{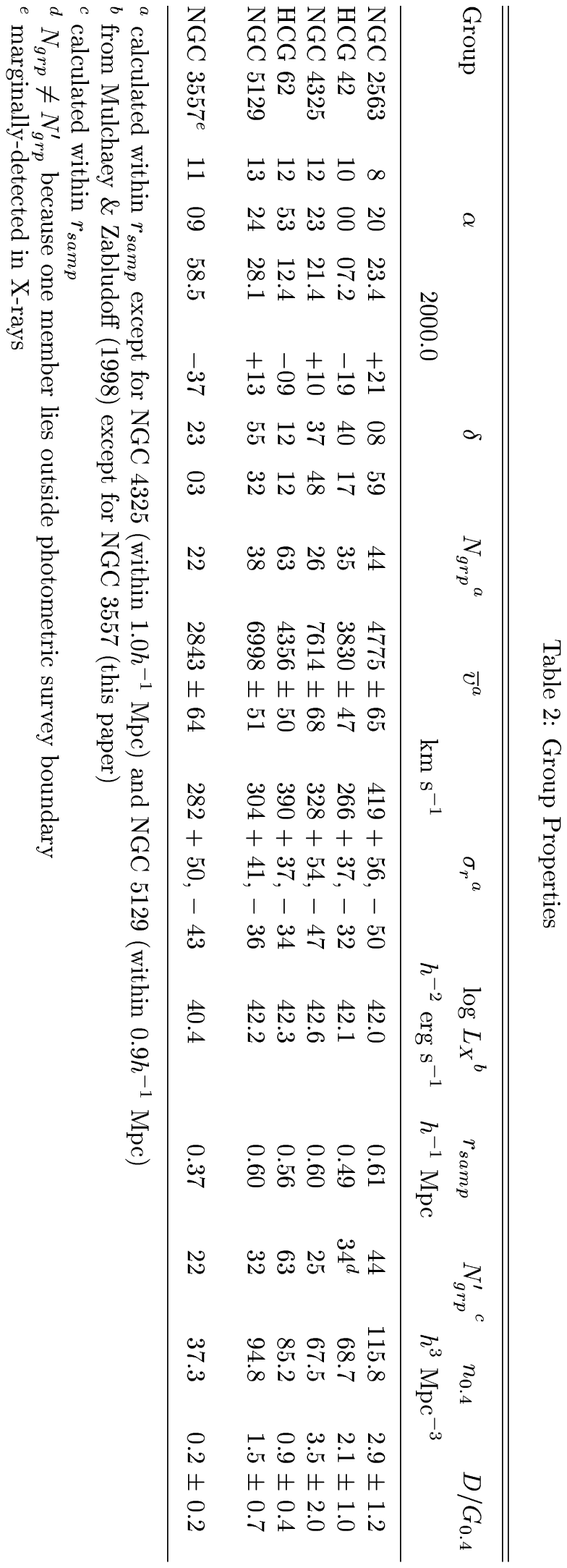}
\end{figure}
\clearpage
\vfill\eject
\centerline{\bf References}
\bigskip
\annrev{Barnes, J.E. \& Hernquist, L. 1992}{30}{705}
\aa{Bertin, E. \& Arnouts, S. 1996}{117}{393}
\annrev{Binggeli, B., Sandage, A., \& Tammann, G. 1988}{26}{509}
\refbook{Blitz, L., Spergel, D., Teuben, P., Hartmann, D., \& Burton, W.B. 1998, preprint (astro-ph/9803251)}
\apj{Bromley, B., Press, W., Lin, H., Kirshner, R. 1998}{505}{25}
\refbook{Colbert, J., Mulchaey, J., \& Zabludoff, A. 1999, in prep.}
\apj{David, L. \& Blumenthal, G. 1992}{389}{510}
\apj{Dressler, A. 1980}{236}{351}
\mnras{Driver, S., Couch, S., \& Phillipps, S. 1998}{287}{415}
\aj{Ferguson, H. \& Sandage, A. 1991}{101}{765}
\aj{Geller, M., Kurtz, M., Wegner, G., Thorstensen, J., Fabricant, D., Marzke, R.,
Huchra, J., Schild, R., \& Falco, E. 1997}{114}{2205}
\pasp{Graham, J. 1982}{94}{244}
\refbook{Grebel, E. 1999, in ``The Stellar Content of the Local Group'', 
IAU Symp. 192, eds. P. Whitelock \& R. Cannon, ASP Conf. Ser.}
\refbook{Helou, G., Madore, G., Schmitz, M.,
Bicay, M., Wu, X. \& Bennett, J. 1991, in ``Databases and On-Line Data
in Astronomy," ed. D. Egret \& M. Albrecht (Dordrecht: Kluwer), p. 89.}
\annrev{Hickson, P. 1997}{35}{357}
\apj{Hickson, P. 1982}{255}{382}
\apj{Hunsberger, S., Charlton, J., \& Zaritsky, D. 1998}{505}{536}
\apj{Hunsberger, S., Charlton, J., \& Zaritsky, D. 1996}{462}{50}
\mnras{James, P. 1991}{250}{544}
\aj{Jarvis, J.F. \& Tyson, J.A. 1981}{86}{476}
\aa{Jerjen, H. \& Tammann, G. 1997}{321}{713}
\mnras{Kauffmann, G., Nusser, A., \& Steinmetz, M. 1997}{286}{795}
\aj{Koranyi, D., Geller, M., Mohr, J., \& Wegner, G. 1998}{116}{2108}
\refbook{Kravtsov, A. \& Klypin, A. 1999, preprint (astro-ph/9812311)}
\apj{Lin, H., Kirshner, R.P., Shectman, S.A., Landy, S.D.,
Oemler, A., Tucker, D. L., Schechter, P. L. 1996}{464}{60}
\refbook{Lin, H. 1995, Ph.D. Thesis, Harvard University}
\apjlett{L\'opez-Cruz, O., Yee, H., Brown, J., Jones, C., \& Forman, W. 1997}{475}{L97}
\mnras{Lorrimer, S., Frenk, C., Smith, R., White, S., \& Zaritsky, D. 1994}{269}{696}
\apj{Loveday, J., Peterson, B., Efstathiou, G., \& Maddox, S.J. 1992}{390}{338}
\aj{Marzke, R., Geller, M., Huchra, J., \& Corwin, H. 1994a}{108}{2}
\apj{Marzke, R., Huchra, J., \& Geller, M. 1994b}{428}{43}
\refbook{Mink, D. J. \& Wyatt, W. F. 1995, Astronomical Data Analysis 
Software and Systems IV, ASP Conference Series, Vol. 77,
1995, R.A. Shaw, H.E. Payne, and J.J.E. Hayes, eds., p. 496.}
\apj{Moore, B., Lake, G., \& Katz, N. 1998}{495}{139}
\nature{Moore, B., Katz, N., Lake, G., Dressler, A., \& Oemler, A. 1996}{379}{613}
\apj{Mulchaey, J. S. \& Zabludoff, A.I. 1998}{496}{73} (MZ98)
\apj{Muriel, H., Valotto, C., \& Lambas, D. 1998}{506}{540}
\refbook{Narayanan, V., Berlind, A., Weinberg, D. 1998, preprint (astro-ph/9812002)}
\apj{Ramella, M., Geller, M.J., \& Huchra, J.P. 1989}{344}{57}
\apj{Schechter, P.L. 1976}{203}{297}
\aj{Schild, R. 1977}{82}{337}
\apj{Secker, J. \& Harris, W. 1996}{469}{623}
\refbook{Shectman, S.A., Schechter, P.L., Oemler, A.A., Tucker, D.,
Kirshner, R.P., \& Lin, H. 1992, in Clusters and Superclusters
of Galaxies (ed. Fabian, A.C.) 351-363 (Kluwer, Dordrecht)}
\apj{Shectman, S.A., Landy, S.D., Oemler, A., Tucker, D.L., Lin, H.,
Kirshner, R.P.; Schechter, P.L. 1996}{470}{172}
\mnras{Smith, R., Driver, S., \& Phillipps, S. 1997}{287}{415}
\aj{Thompson, L. \& Gregory, L. 1993}{106}{2197}
\mnras{Trentham, N. 1997}{290}{334}
\refbook{Tucker, D., Hashimoto, Y., Kirshner, R., Landy, S., Lin, H.,
Oemler, A., Schechter, P., \& Shectman, S. 1998,
in Large Scale Structure: Tracks and Traces, proceedings of the 12th Potsdam Cosmology Workshop
(ed. V. Mueller, S. Gottloeber, J.P. Muecket, J. Wambsganss) 105-106 (World Scientific)}
\apj{Valotto, C. A., Nicotra, M. A., Muriel, H., \& Lambas, D. G. 1997}{479}{90}
\nature{White, S., Davis, M., Efstathiou, G., \& Frenk, C. 1987}{330}{451}
\mnras{Wilson, G., Smail, I., Ellis, R., \& Couch, W. 1997}{284}{915}
\apj{Zabludoff, A.I. \& Mulchaey, J.S. 1998}{496}{39} (ZM98)
\aj{Zaritsky, D., Zabludoff, A.I., \& Willick, J. 1995}{110}{1602}
\apj{Zepf, S., de Carvalho, R., \& Ribeiro, A. 1997}{488}{11}

\vfill\eject\noindent

\centerline{\bf Figure Captions}

\bigskip
\noindent
{\bf Figure 1:}  Contour map of the diffuse X-ray emission 
in the NGC 3557 group of
galaxies overlaid on the STScI Digital Sky Survey.  
The X-ray point sources have been removed following the procedure outlined
in MZ98.  The contours
correspond to levels 3, 4, and $5\sigma$ above the background. The
data have been smoothed with a Gaussian profile of width 30$''$. The
coordinate axes are J2000.

\noindent
{\bf Figure 2:} Galaxy velocity distributions out to 30000 \ks\ for
the six poor groups in our sample.  The shaded histograms indicate the
N$_{grp}$ group members identified with the pessimistic $3\sigma$-clipping
algorithm described in ZM98.  In the group HCG 42, 
we manually add one galaxy (H42\_136; $\upsilon = 4587$ \ks) excluded by
the membership algorithm 
to the membership list, because this galaxy's velocity 
bin is contiguous with the group's velocity peak.

\noindent
{\bf Figure 3:}  Distributions of apparent R magnitudes ($m_R$) for galaxies with
measured velocities in the six poor group fields.  The number
above each bar indicates the percentage of the total number of galaxies
in the field within that magnitude bin represented by the plotted
galaxies.  Differences between N$_{tot}$ in Figure
2 and N$_{spec}$ here are due to galaxies observed spectroscopically
that lie just off the edge of the photometric field.
In two groups, NGC 4325 and NGC 5129, our spectroscopy and
imaging extend beyond the radius of 0.6\lith\inv\ Mpc sampled in the
other groups.  In these two cases, we also show the $m_R$ distribution
for the subset of galaxies within 0.6\lith\inv\ Mpc that is used for
all subsequent analyses in this paper (shaded).

\noindent
{\bf Figure 4:} Galaxy luminosity distributions for the members of
each of the six groups in the sample.  The first five groups, HCG 42,
HCG 62, NGC 2563, NGC 4325, and NGC 5129 are X-ray luminous, whereas
the last group, NGC 3557, is only marginally X-ray detected.  The
total number of spectroscopically-confirmed group members is
${N}^{\prime}_{grp}$.  The
absolute magnitudes $M_R$ are extinction-corrected and calculated
assuming a $H_0 = 100$ \ks\ Mpc$^{-1}$, $q_0 = 0.5$ cosmology.  
The shaded boxes
are the observed number of group members within that
magnitude bin.  The solid boxes are the completeness-corrected galaxy
counts ($\S3.1$).  (For HCG 62, NGC 4325, and NGC 3557, the
corrected counts exceed the limit of the y-axis at the faintest magnitudes.)

\noindent
{\bf Figure 5:}  (a)  Top panel:  Comparison of the galaxy 
luminosity function for the composite of the five X-ray
luminous groups (filled triangles) and for two composites
of nearby rich clusters
of galaxies (short dashed line, Trentham 1997; dot-dashed line,
Driver \etal 1998).  
To simplify the comparison of the GLF shapes,
the curves in panels (a) and (b)
are normalized to have the same total number of $M_R
\leq -17 + 5$log $h$ galaxies as HCG 42.  
The composite group GLF is derived from averaging the
completeness-corrected counts in Figure 4 after normalizing the
individual group GLF's to the same total number of $M_R \leq -17 +
5$log $h$ galaxies as HCG 42.  The best fit to the group GLF is
consistent with a Schechter function with $\Mstar = -21.6 \pm 0.4 +
5$log $h$ and $\alpha = -1.3 \pm 0.1$ (thick solid line in both panels).
The three composite GLF's are indistinguishable for the given errors.
(b)  Bottom panel:  Comparison of the group GLF
in (a) with that of the
Campanas Redshift Survey (LCRS) field (long dashed line; Lin \etal 1996)
to the completeness limit of the LCRS ($M_R \sim -17.5 + 5$log $h$).
The LCRS and composite group GLF's differ
at the $> 95\%$ confidence level for any choice of relative normalization.  
A flat faint end slope of $\alpha = -1$
is also plotted for comparison.

\noindent
{\bf Figure 6:} Comparison of the GLF for star forming and for
quiescent galaxies in X-ray groups and in the LCRS field.  The
composite group GLF in Figure 5 is split here into (1) the GLF for
galaxies whose spectra have [OII] EW $\geq 5$ \AA\ 
(open triangles) and (2) the GLF for galaxies with [OII] EW $< 5$ \AA\ 
(filled circles).  The GLF for the
LCRS field is split similarly into star forming (short dashed line)
and quiescent (long dashed line) components.  Once again each
component is arbitrarily 
normalized to the to the same total number of $M_R \leq
-17 + 5$log $h$ star forming or quiescent galaxies as HCG 42.  The thick
solid line is as in Figure 5b.

\noindent
{\bf Figure 7:}  (a) Left panel: Velocity offset vs. projected 
radial offset of 123
quiescent group members from the group centroid for the six groups in
the sample.  The velocity offset is normalized with the group velocity
dispersion ($\sigma_{grp}$).  The six asterisks are 
four of the brightest group
galaxies (BGGs) and two other galaxies
with $M_R < \Mstar$.  The open circles
are the 56 giants defined by $\Mstar \leq M_R \leq -19 + 5$log $h$.  
The filled circles are the 61 dwarfs defined by $-19 + 5$log $h
< M_R \leq -17 + 5$log $h$.  (Note that the data extend to a
projected radius of $\simgreat 0.6$\lith\inv\ Mpc $> r_{samp}$,
because the group centroid shown here and the fiber
field center are not precisely coincident in some cases.)  Right
panel: The distribution of $R$ ($\S3.4$), the quadrature sum of the
x- and y-axis offsets of each galaxy, for the BGG (heavily shaded),
giant (shaded), and dwarf populations (unshaded).  The $R$
distributions suggest that the three populations occupy different
orbits ({\it i.e.}, have not mixed completely).  (b) The same as in
(a) for 49 star forming group members.

\noindent
{\bf Figure 8:}  $D/G$ profile for
the quiescent members of each group in Figure 7a. 
The significance of the correlation as determined from a Spearman rank-order 
test is $> 95\%$.
The trend is likely to be even steeper than shown, because
the sample includes two Hickson Compact Groups, 
which have unusually low core $D/G$ values 
(the two lowest filled circles in the first bin) due to the
Hickson Group selection criteria.
Removing the marginally X-ray detected group NGC 3557 (open circles), 
which is sampled only to 0.4\lith\inv\ Mpc and has the lowest $D/G_{0.4}$,
increases the steepness of the trend and the significance of the 
Spearman correlation coefficient.

\eject
\clearpage
\setcounter{figure}{0}
\begin{figure}
\plotone{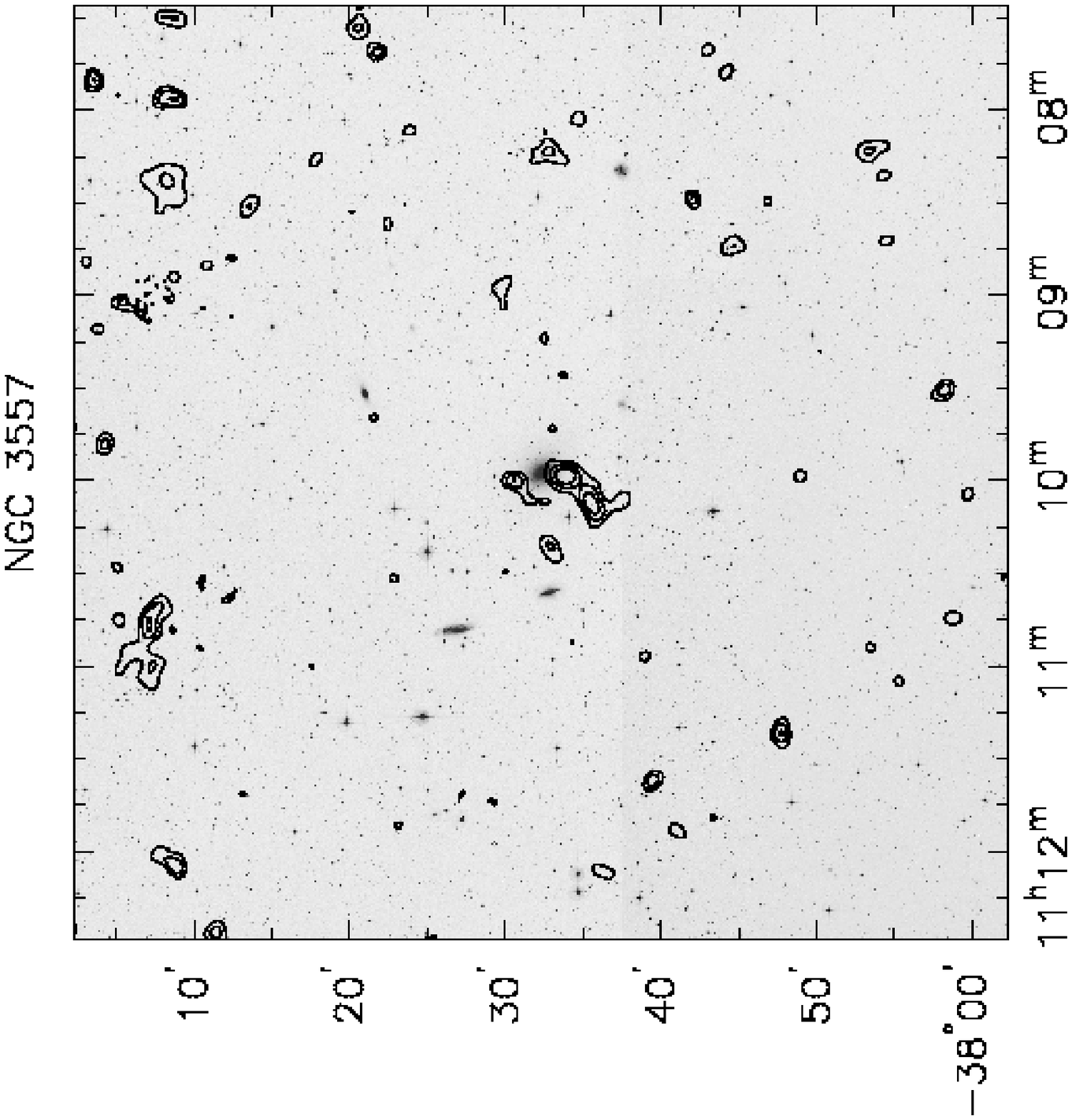}
\caption{}
\end{figure}
\clearpage
\vfill\eject
\clearpage
\setcounter{figure}{1}
\begin{figure}
\plotone{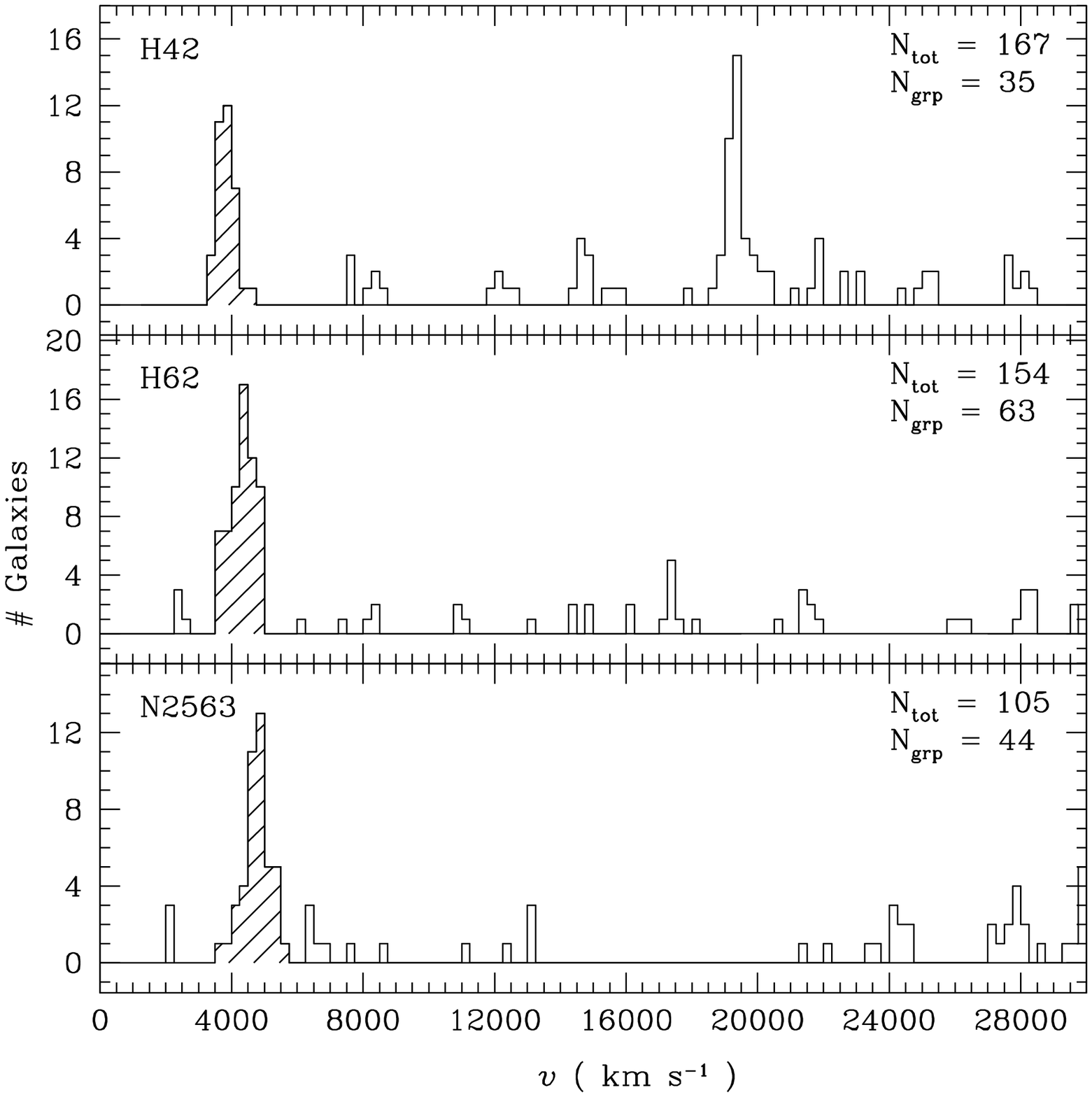}
\caption{}
\end{figure}
\clearpage
\vfill\eject
\clearpage
\setcounter{figure}{1}
\begin{figure}
\plotone{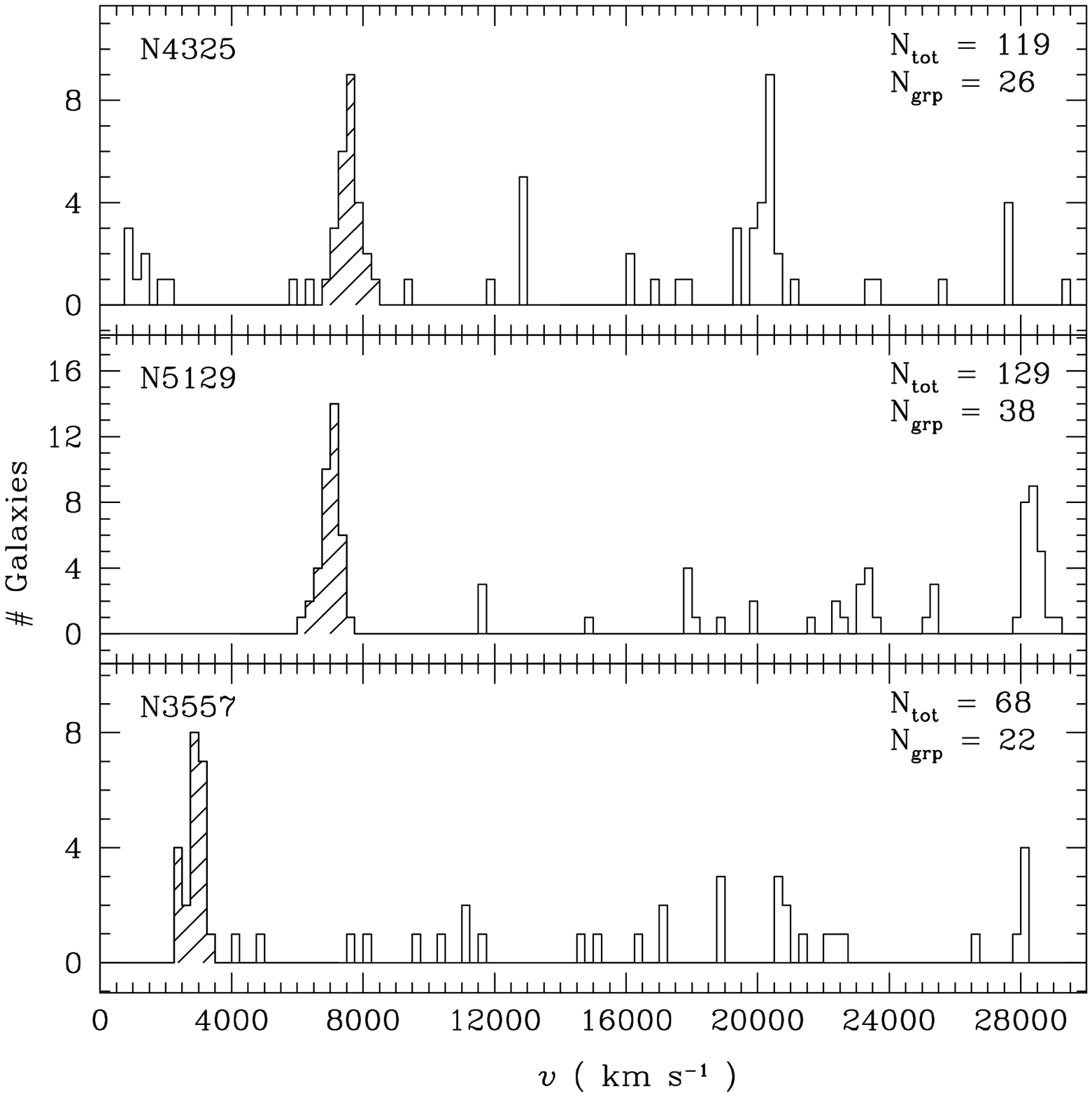}
\caption{{\it cont. }}
\end{figure}
\clearpage
\vfill\eject
\clearpage
\setcounter{figure}{2}
\begin{figure}
\plotone{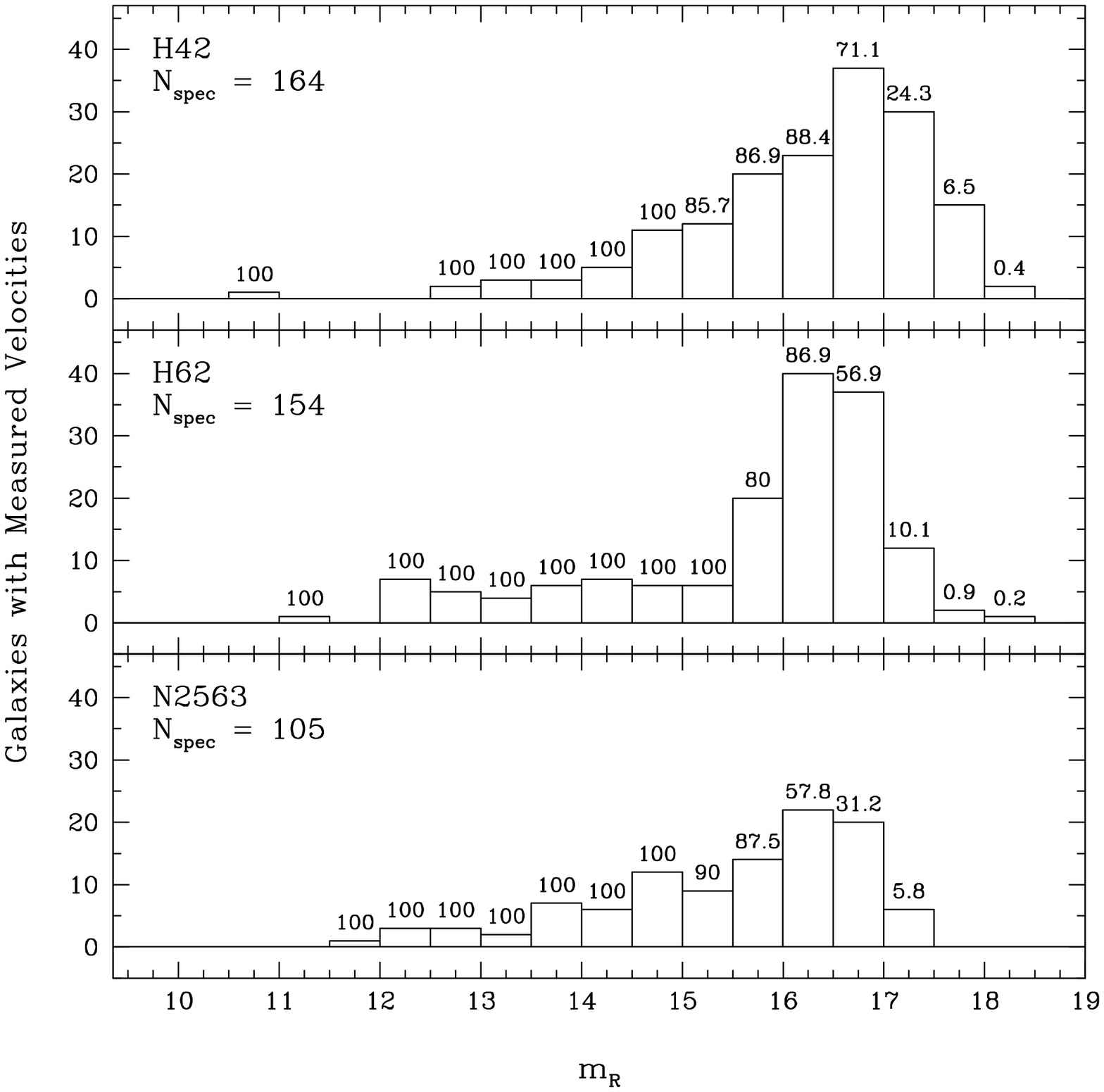}
\caption{}
\end{figure}
\clearpage
\vfill\eject
\clearpage
\setcounter{figure}{2}
\begin{figure}
\plotone{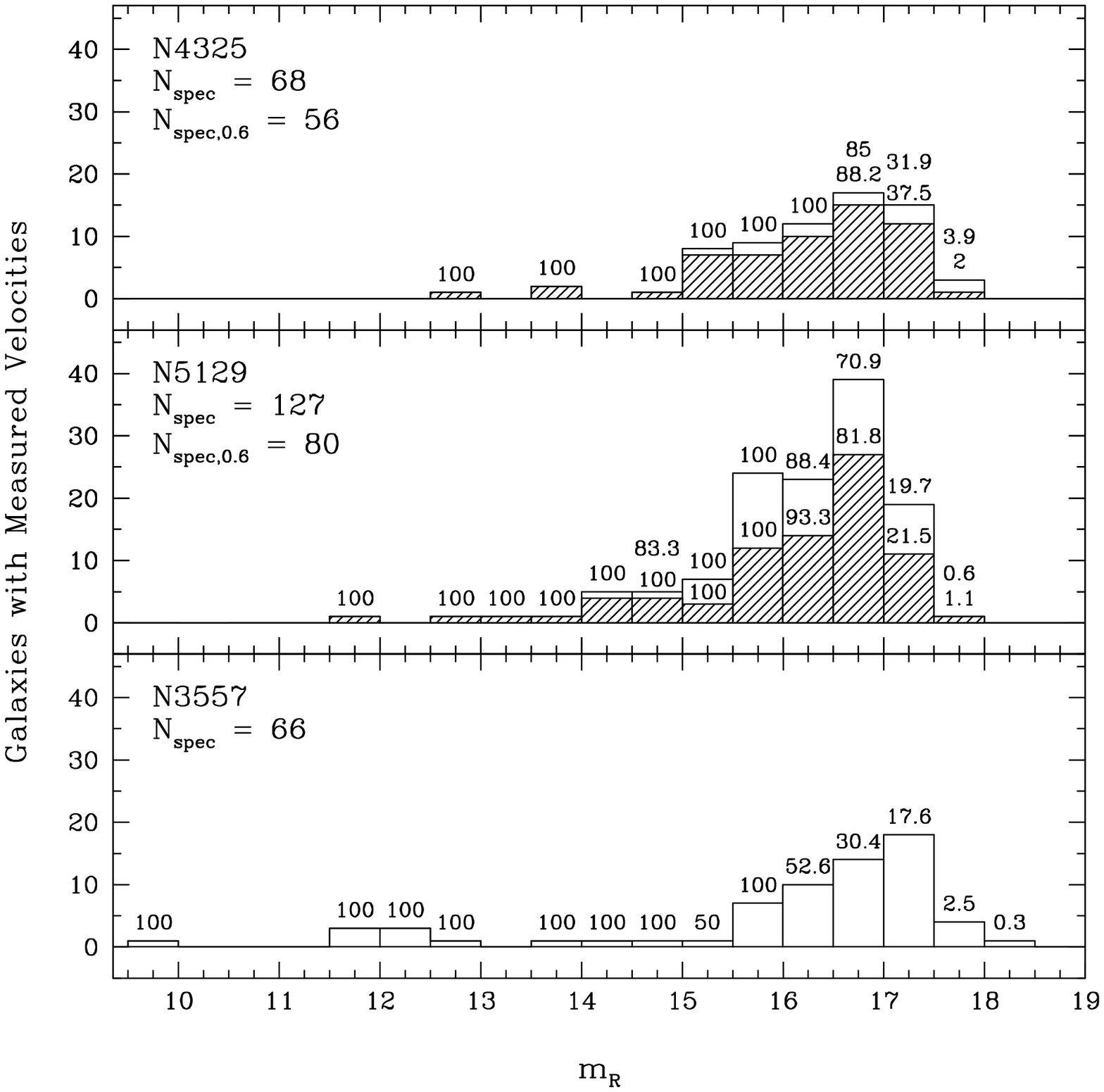}
\caption{{\it cont. }}
\end{figure}
\clearpage
\vfill\eject
\clearpage
\setcounter{figure}{3}
\begin{figure}
\plotone{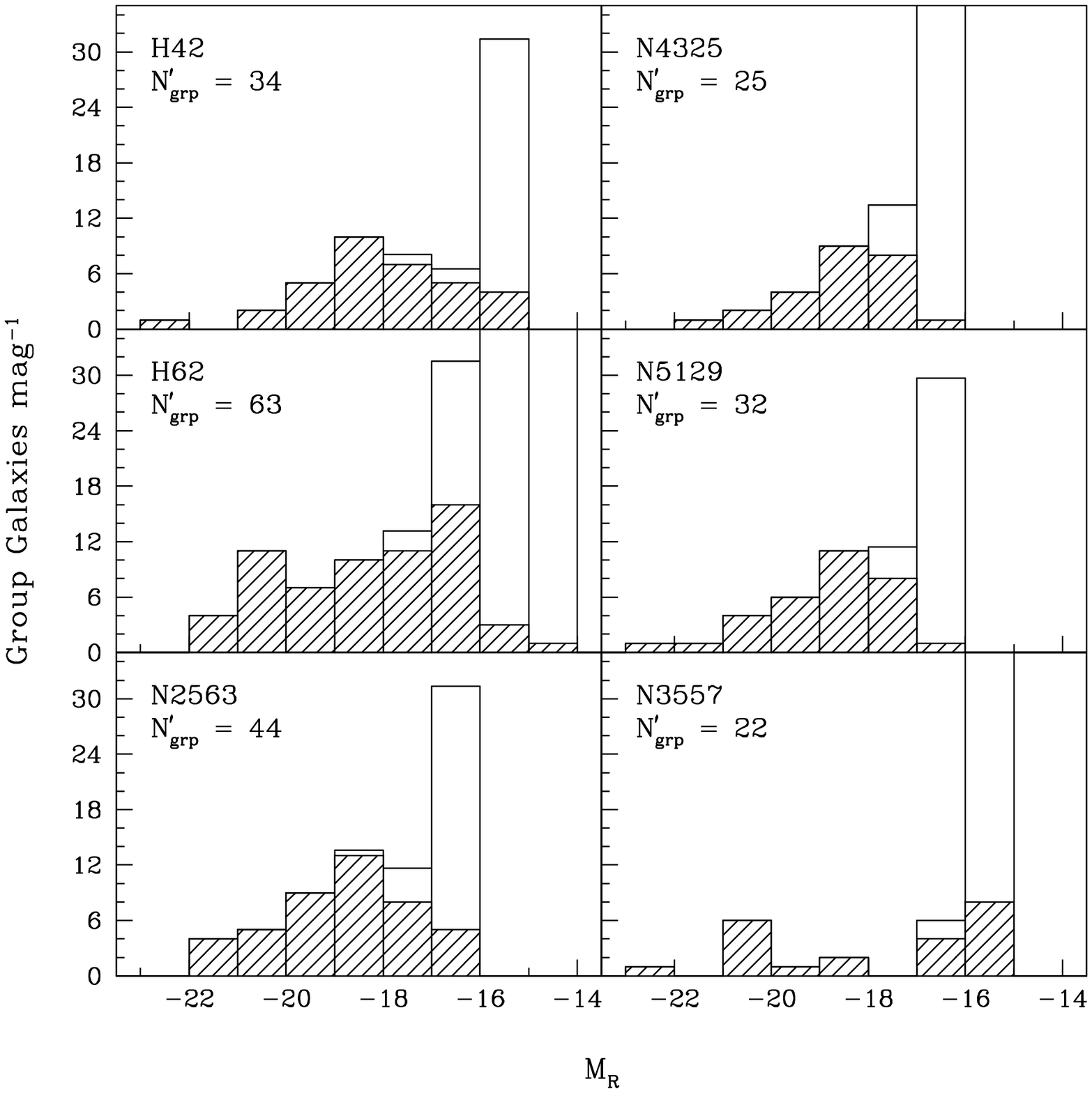}
\caption{}
\end{figure}
\clearpage
\vfill\eject
\clearpage
\setcounter{figure}{4}
\begin{figure}
\plotone{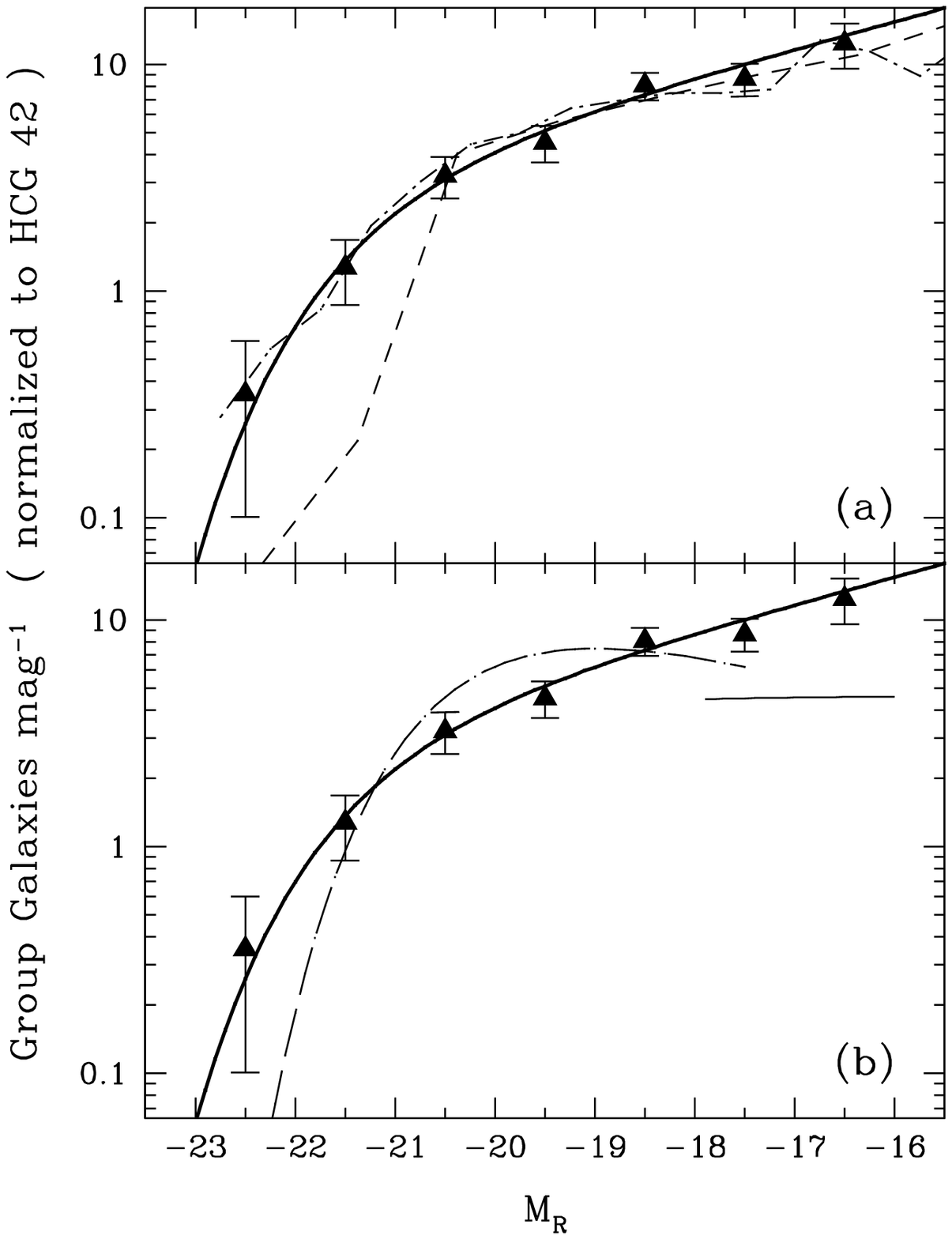}
\caption{}
\end{figure}
\clearpage
\vfill\eject
\clearpage
\begin{figure}
\plotone{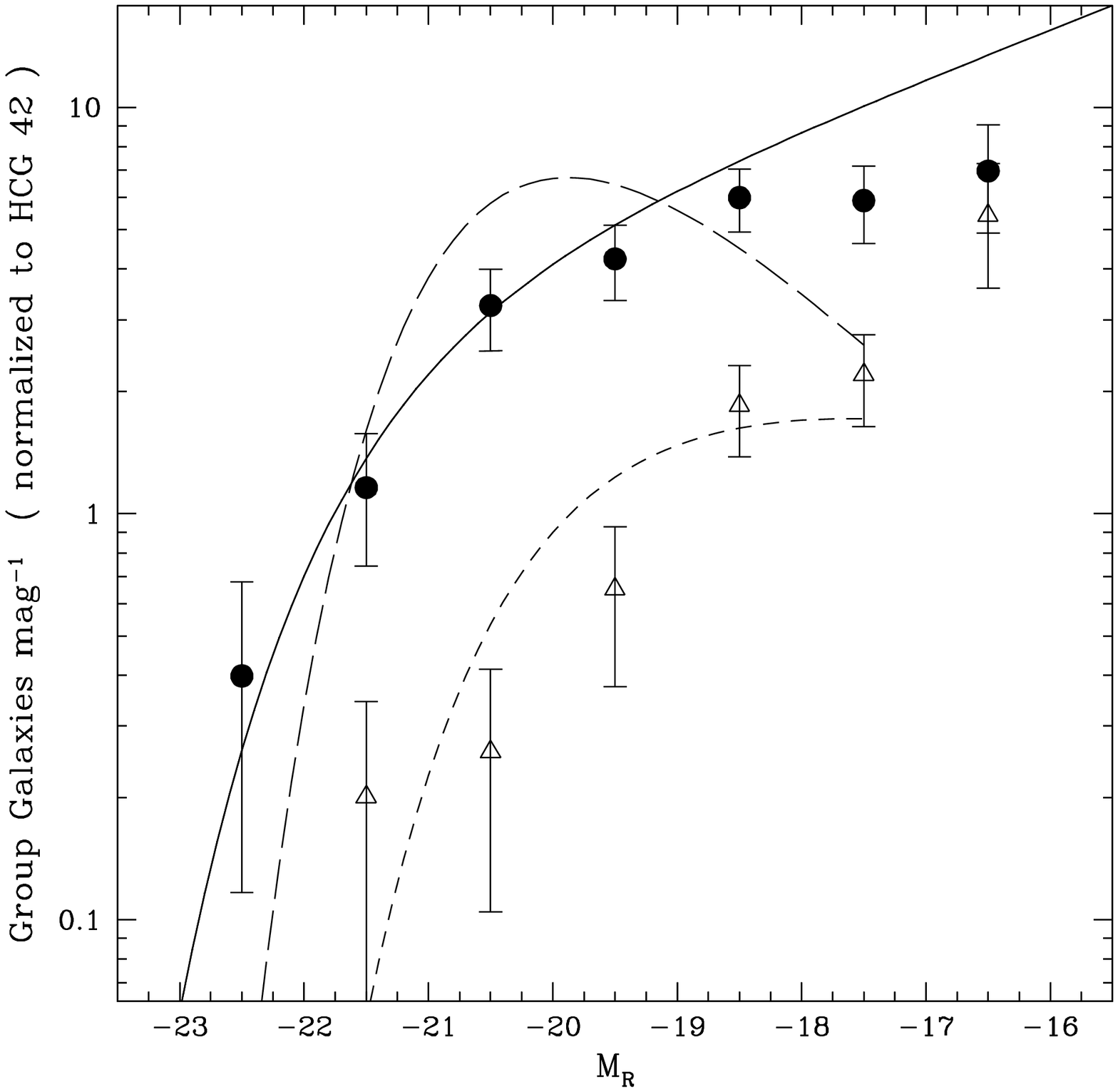}
\caption{}
\end{figure}
\clearpage
\vfill\eject
\clearpage
\begin{figure}
\plotone{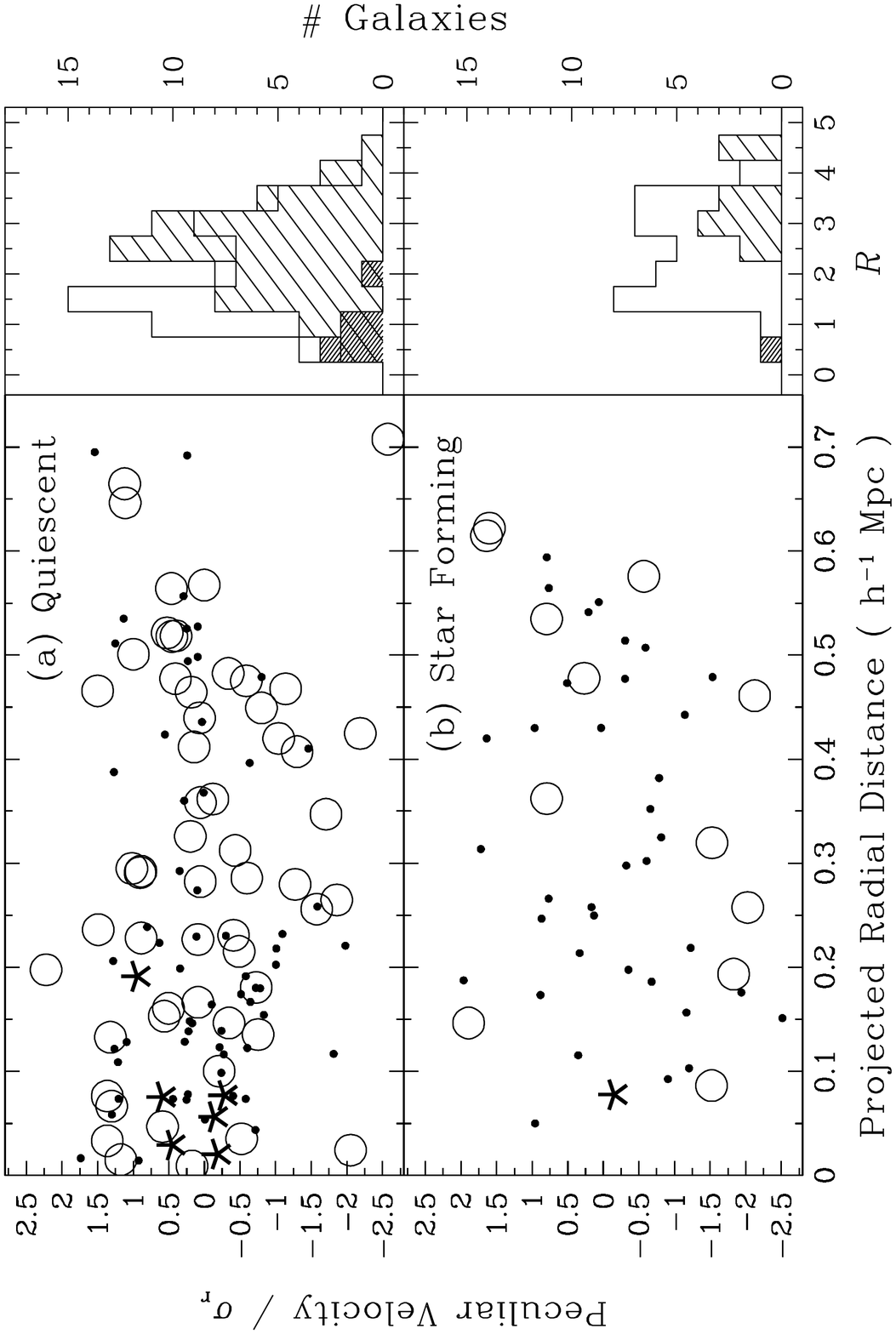}
\caption{}
\end{figure}
\clearpage
\vfill\eject
\clearpage
\begin{figure}
\plotone{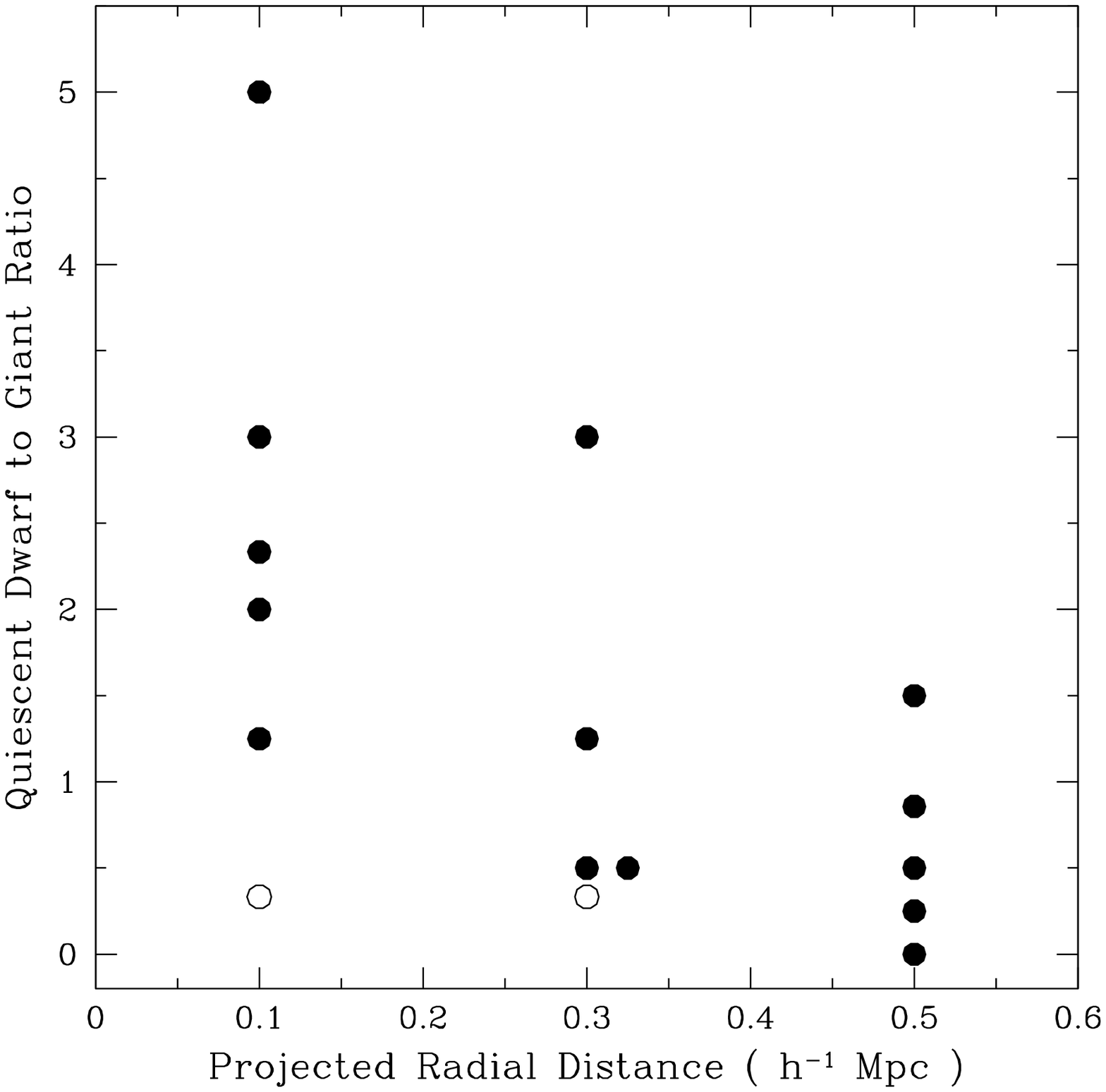}
\caption{}
\end{figure}
\clearpage
\vfill\eject
\end{document}